\documentclass[prl,a4paper,11pt]{article}
\pdfoutput=1

\usepackage{bm}        % Include bold math: \bm{} creates bold letters symbols in math mode
\usepackage{amsmath}						%math symbols%
\usepackage{amssymb}						%additional math symbols%
\usepackage{amsthm}						%math theorems%
\usepackage{array}							%create more sophisticated tables%
\usepackage[english]{babel}					%language specific characters, hyphernation, etc.%
\usepackage{cancel}						%cross out symbols%
\usepackage{eufrak}							%additional font%
\usepackage{fullpage}						%narrower margins%
\usepackage{graphicx}						%include graphics%
\usepackage{multirow}						%cell over multiple rows in table%
\usepackage{nccmath}						%even more additional math symbols%
\usepackage{nicefrac}						%nice fracs%
\usepackage[section]{placeins}                         %in the preamble, plots, tables forced to remain in the section
\usepackage{booktabs}						%prettier tables%
\usepackage[format=hang]{caption}				%hang option for mulit-line captions%
\usepackage[hang,flushmargin]{footmisc}		%hang option for footnotes%
\usepackage{hyperref}						%include hyperlinks%
\usepackage{mathtools}						% extension and bug fixing form amsmath package%	

\numberwithin{equation}{section}				%number equations by section%
\newcommand{\balpha}{\bar{\alpha}}			%Greek alphabet \bar%
\newcommand{\bbeta}{\bar{\beta}}
\newcommand{\bgamma}{\bar{\gamma}}
\newcommand{\bdelta}{\bar{\delta}}
\newcommand{\bepsilon}{\bar{\epsilon}}
\newcommand{\bmu}{\bar{\mu}}
\newcommand{\bpi}{\bar{\pi}}
\newcommand{\brho}{\bar{\rho}}
\newcommand{\btau}{\bar{\tau}}

\newcommand{\NPl}{\bm{\ell}}					%bold letters for NP tetrad%
\newcommand{\NPn}{\bm{n}}
\newcommand{\NPm}{\bm{m}}
\newcommand{\NPbm}{\bm{\overline{m}}}

\def\scalar{{\mathbf S}} 
\def\vector{{\mathbf V}}

\begin{document}

\thispagestyle{empty}

\vspace*{2cm}

\begin{center}
%title%
{\bf \LARGE Boundary Conditions for Kerr-AdS Perturbations}\\
\vspace*{2.5cm}

%authors%
{\bf \'Oscar J. C.~Dias$^\star$, } {\bf Jorge E.~Santos$^\ddagger$}

\vspace*{1cm}

%affiliations%
 {\it $\star$ Institut de Physique Th\'eorique, CEA Saclay,\\
 CNRS URA 2306, F-91191 Gif-sur-Yvette, France}\\ \vspace{0.3cm}
 {\it $\ddagger$ Department of Physics, UCSB, Santa Barbara, CA 93106, USA}\\ 
 \vspace{0.3cm}

\vspace*{0.5cm} {\tt oscar.dias@cea.fr, jss55@physics.ucsb.edu}

%\vspace{0.5cm} {date}
\end{center}

\vspace*{1cm}

%abstract%
\begin{abstract}
The Teukolsky master equation and its associated spin-weighted spheroidal harmonic decomposition simplify considerably the study of linear gravitational perturbations of the Kerr(-AdS) black hole. However, the formulation of the problem is not complete before we assign the physically relevant boundary conditions. We find a set of two Robin boundary conditions (BCs) that must be imposed on the Teukolsky master variables to get perturbations that are asymptotically global AdS, i.e.  that  asymptotes to the Einstein Static Universe.  
In the context of the AdS/CFT correspondence, these BCs allow a non-zero expectation value for the CFT stress-energy tensor while keeping fixed the boundary metric.
When the rotation vanishes, we also find the gauge invariant differential map between the Teukolsky and the Kodama-Ishisbashi (Regge-Wheeler$-$Zerilli) formalisms. One of our Robin BCs maps to the scalar sector and the other to the vector sector of the Kodama-Ishisbashi decomposition.
The Robin BCs on the Teukolsky variables will allow for a quantitative study of instability timescales and quasinormal mode spectrum of the Kerr-AdS black hole.
As a warm-up for this programme, we use the Teukolsky formalism to recover the quasinormal mode spectrum of global AdS-Schwarzschild, complementing previous analysis in the literature.
\end{abstract}
\noindent

\newpage
%table of contents%
\thispagestyle{empty}
\tableofcontents

%set counters to zero%
\setcounter{page}{0} \setcounter{footnote}{0}

%%%%%%%%%%%%%%%%%%%%%%%%%%%%%%%%%%%%%%%%%%%%%%%%%%%%
\section{Introduction and summary\label{sec:IntroSum}}
%%%%%%%%%%%%%%%%%%%%%%%%%%

It is unquestionable that few systems are isolated in Nature and we can learn a lot from studying their interactions. Black holes are no exception and the study of their perturbations and interactions reveals their properties (see e.g. the recent roadmap \cite{Cardoso:2012qm} and review \cite{Berti:2009kk} on the subject). The simplest  deformation we can introduce in a background is a linear perturbation, which often encodes interesting physics such as linear stability of the system and its quasinormal mode spectrum. Moreover, it also anticipates some non-linear level properties. For example, in the collision of two black holes, such as in the coalescence of a binary system, after the inspiral and merger phase, the system undergoes a ring down phase where gravitational wave emission is dictated by the quasinormal mode frequencies. The linear perturbation fingerprints are therefore valuable from a theoretical and gravitational-wave detection perspective \cite{Cardoso:2012qm,Berti:2009kk}. Perhaps more surprisingly, linear analysis of black holes in AdS can be used to infer properties about their nonlinear stability \cite{Bizon:2011gg,Dias:2011ss,Dias:2012tq}. Linear analysis can also infer some properties of (nonlinear) black hole collisions and associated gravitational wave emission in the close-limit approximation \cite{Price:1994pm}.

To study linear gravitational perturbations of a black hole we need to solve the linearized Einstein equation. {\it \`A priori} this is a remarkable task involving a coupled  system of PDEs. Fortunately, for the Kerr(-AdS) black holes (which are Petrov type D backgrounds), Teukolsky employed the Newman-Penrose formalism to prove that all the gravitational perturbation information is encoded in two decoupled complex Weyl scalars \cite{Teukolsky:1972my,Teukolsky:1973ha}. These are gauge invariant quantities with the same number of degrees of freedom as the metric perturbation. Moreover, there is a single pair of decoupled master equations governing the perturbations of these Weyl scalars (one element of the pair describes  spin $s=2$ and the other $s=-2$ modes). In a mode by mode analysis, each master equation further separates into a radial and angular equation which makes the analysis technically tractable \cite{Teukolsky:1973ha,Chandra1,Chandra2,ChandrasekharBook}. (In the absence of rotation, and only in this case, we can instead use a similar pair of decoupled master equations for a distinct pair of gauge invariant variables proposed by Regge and Wheeler \cite{Regge:1957td} and Zerilli  \cite{Zerilli:1970se}, and later recovered and extended by Kodama and Ishibashi \cite{Kodama:2003jz}). 

Solving these master equations is not our only task. Like in any PDE system, it is also important to assign physically relevant boundary conditions. Without the later, the formulation of the problem is not complete.

In this paper we are interested in linear gravitational perturbations of the Kerr-AdS black hole, with a focus on its boundary conditions (BCs). The (extra) motivation to put into a firmer basis the linear perturbation problem of the Kerr-AdS system is two-folded. First, the Kerr-AdS black hole is known to have linear gravitational instabilities sourced by superradiance \cite{Hawking:1999dp}-\cite{Cardoso:2006wa} and by extremality \cite{Lucietti:2012sf}. Second, in the AdS/CFT duality context,  perturbing a (Kerr-AdS) black hole in the bulk is dual to perturbing the associated CFT thermal state (with a chemical potential) living in bulk boundary. The time evolution of the perturbed black hole maps into the  time evolution of the thermal state fluctuations and the quasinormal mode spectrum of the black hole is dual to the thermalization timescale in the CFT (see e.g   \cite{Horowitz:1999jd}-\cite{Friess:2006kw}, \cite{Berti:2009kk}). 

From the mathematical perspective, the boundary condition choice is arbitrary. We need physical input to fix it. Not always but quite often, this leads to a unique choice.  
 We establish what are the BCs we need to impose in the Teukolsky master solutions to get perturbations that are {\it asymptotically global AdS}. To make this statement precise, recall that once we have the solution  of the Teukolsky pair of master variables  ($s=\pm 2$), we can reconstruct the metric perturbations using the Hertz map \cite{Cohen:1975}-\cite{Dias:2009ex}. We get a pair of metric perturbations, one in the ingoing radiation gauge (IRG; $s=-2$) and the other in the outgoing radiation gauge (ORG; $s=2$). By asymptotically global AdS perturbations we mean that we want {\it the} BCs in the Teukolsky scalars that yield metric perturbations that decay at asymptotic infinity according to  {\it the}  power law found by Henneaux and Teiltelboim \cite{Boucher:1983cv,Henneaux:1985tv}. Our task is thus very well defined. We have to work out the inverse Hertz map and find how the Henneaux-Teiltelboim metric BCs translate into the Teukolsky scalars.

Before arguing further that this choice should be the  physically relevant option, it is illuminating to recall what is the situation in an asymptotically flat system. In this case, the BC choice in the Teukolsky scalars amounts to choosing the purely outgoing traveling mode. Intuitively, this is because we are not interested in scattering experiments (where an ingoing mode component would be present). Formally, this is because this is the  choice that yields a metric perturbation preserving the asymptotic flatness of the original Kerr black hole, i.e. conserving asymptotically the Poincar\'e group of the Minkowski spacetime. 

A similarly reasoning justifies why the Henneaux-Teiltelboim BCs should be the physically relevant boundary condition to be imposed in the Kerr-AdS system \cite{Henneaux:1985tv}. These are the BCs that  preserve asymptotically  the global AdS symmetry group $O (3, 2)$ and yield finite surface integral charges associated with the  $O(3, 2)$ generators. Yet an additional reason to single out this BC is justified by the AdS/CFT duality.
The Kerr-AdS is asymptotically global AdS and the CFT lives on the boundary of this bulk spacetime. As desired in this context, the Henneaux-Teiltelboim BCs are such that they allow for a non-zero expectation value for the CFT stress-energy tensor while keeping fixed the boundary metric. This criterion to select the BCs of gauge invariant variables was first emphasized in the context of the Kodama-Ishibashi (KI) formalism \cite{Kodama:2003jz} by Michalogiorgakis and Pufu \cite{Michalogiorgakis:2006jc}. They pointed out that  previous analysis of quasinormal modes of the 4-dimensional global AdS-Schwarzschild black hole using the KI master equations were preserving the boundary metric only in the KI vector sector, but not in the KI scalar sector of perturbations (here, vector/sector refer to the standard KI classification of perturbations). Indeed, previous studies in the literature had been imposing Dirichlet BCs on the KI gauge invariant variables. It turns out that in the KI scalar sector keeping the boundary metric fixed requires a Robin BC (which relates the field with its derivate) \cite{Michalogiorgakis:2006jc}.    
Still in the context of AdS/CFT on a sphere, other boundary conditions that might be called asymptotically globally AdS (and that promote the boundary graviton to a dynamical field) were proposed in \cite{Compere:2008us}. However, they turn out to lead to ghosts (modes with negative kinetic energy) and thus make the energy unbounded below \cite{Andrade:2011dg}. So, the Henneaux-Teiltelboim BCs are also the physically relevant BCs for the AdS/CFT where the CFT lives in the Einstein Static Universe. 

So, a global AdS geometry with Henneaux-Teitelboim BCs does not deform the boundary metric. This is the mathematical statement materializing the  pictoric  idea that a global AdS background behaves like a confining box with a reflecting wall. An interesting observation that emerges from our study is that these BCs require that we consider a particular linear combination of the Teukolsky IRG and ORG metric contributions. We can interpret this property as being a manifestation of the common lore that only a standing wave with a node at the AdS boundary can fit inside the confining box.
This pictorial  notion of a standing wave and node is very appealing but, what is the formal definition of a node in the present context? Does it mean that we have to impose a Dirichlet BC on the Teukolsky scalars? No. Instead we will find the Robin BC  \eqref{TheBCs}-\eqref{TheBCaux}, much like what happens in the scalar sector of the aforementioned 4-dimensional KI system. An inspection of this Robin BC (pair) immediately convinces us that we hardly could guess it without the actual computation.

At  first sight the fact the asymptotically global AdS BC requires a sum of the Teukolsky IRG and ORG metric components is rather surprising and, even worrying. Surprising because in the asymptotically flat case we just need to use the outgoing contribution. Eventually worrying because it is known that in Petrov type D backgrounds the two Teukolsky families of perturbations ($s=\pm 2$) encode the same information, once we use the Starobinsky-Teukolsky identities \cite{Staro,StaroChuri,Teukolsky:1974yv,Chandra1,Chandra2} that fix the relative normalization between the $s=\pm 2$ Weyl scalars perturbations. Our result is however not in contradiction with this property. Indeed, the previous statement implies that the most general solution of the $s=\pm 2$ master equations contain the same information but says nothing about the BCs. This just highlights that the differential equations and the BCs are two distinct aspects of the problem, which is not a surprise.  Once we find our BCs, in practical applications, we just need to study the $s=2$ (say) Teukolsky sector of perturbations.  
We believe that  an infinitesimal rotation of the tetrad basis should allow to derive our results using only the outgoing gauge (say), although at the cost of loosing contact with the standing wave picture.

We have already mentioned that perturbations for static backgrounds (with global AdS and global AdS-Schwarzschild being the relevant geometries here) can be studied using the Kodama-Ishibashi (KI) gauge invariant formalism  \cite{Kodama:2003jz} (i.e. the Regge-Wheeler$-$Zerilli formalism \cite{Regge:1957td,Zerilli:1970se}).
On the other hand, the Teukolsky formalism also describes these cases when rotation is absent. 
Therefore, the two formalisms must be related in spite of their differences, although this one-to-one map has not been worked out to date. We fill this gap in the literature. 
The difference that stands out the most  is that the KI formalism decomposes the gravitational perturbations in {\it scalar and vector  spherical} harmonics while  the Teukolsky  formalism uses instead a harmonic decomposition with respect to the {\it spin-weighted spherical} harmonics. These harmonics are distinct and ultimately responsible for the different routes taken by the two formalisms. However, both the KI spherical harmonics and the spin-weighted spherical harmonics can be expressed in terms of the standard scalar spherical harmonic (associated Legendre polynomials) and their derivatives. These two maps establish the necessary bridge between the angular decomposition of the two formalisms. We then need to work out the radial map, which follows from the fact that the metric perturbations of the two formalisms must be the same modulo gauge transformations. This gauge invariant differential map expresses the KI master variables (for the KI scalar and vector sector) in terms of the $s=2$ (say) Teukolsky master field and its first radial derivative and is given in  \eqref{scalarPhiTeuk}-\eqref{vectorPhiTeuk}.
To have the complete map between the KI and Teukolsky ($a=0$) formalisms we also need to discuss the relation between the asymptotically global AdS KI BCs  and the global AdS Teukolsky BCs. This is done in \eqref{TheBCsSa0}-\eqref{vectorPhiTeukBC}. The fact that our BCs for the Teukolsky variables match the 
Michalogiorgakis-Pufu BCs for the KI variables is a non-trivial check of our computation in the limit $a\to 0$.  
Yet this exercise reveals to be more profitable. Indeed, an interesting outcome is that there is a Teukolsky solution/BC that maps to a  KI scalar mode/BC and a second one that maps to a KI vector mode/BC. This is the simplest possible map between the two formalisms that  could have been predicted, yet still a surprise.

With our asymptotically global AdS boundary conditions, the Kerr-AdS linear perturbation problem is completely formulated and ready to be applied to problems of physical interest. These include  finding the quasinormal mode spectrum of Kerr-AdS and the dual CFT thermalization timescales and studying quantitatively the superradiant instability timescale of the solution. This programme is already undergoing and will be presented elsewhere. Neverthless, as a first application, we can recover the quasinormal mode spectrum of the global AdS-Schwarzschild this time using the Teukolsky approach. As it could not be otherwise, we recover the previous results in the literature both for the KI vector sector (firstly studied in \cite{Cardoso:2001bb,Berti:2003ud,Natario:2004jd}) and for the KI scalar sector first obtained in  \cite{Michalogiorgakis:2006jc}.\footnote{\label{foot}As noticed in \cite{Michalogiorgakis:2006jc} the analysis done in  \cite{Cardoso:2001bb}-\cite{Siopsis:2007wn} in the KI scalar case does not impose asymptotically global AdS BCs and thus we will not discuss further the scalar results of these studies.} Our results and presentation contribute to complement these analysis by plotting the spectrum as a function of the horizon radius, and not just a few points of the spectrum. Our analysis focus on the parameter space region of  $r_+/L$ (the horizon radius in AdS units) where the spectrum meets the normal modes of AdS and where it varies the most. We will not discuss the asymptotic (for large overtone  \cite{Natario:2004jd}-\cite{Siopsis:2007wn} and for large harmonic \cite{Dias:2012tq}) behaviour of the QN mode spectrum. 

The plan of this paper is the following. Section \ref{sec:KerrAdS} discusses the Kerr-AdS black hole in the Chambers-Moss coordinate frame \cite{Chambers:1994ap} (instead of the original Carter frame \cite{Carter:1968ks}) that simplifies considerably our future discussion of the results. We discuss the  Teukolsky formalism, the associated Starobinsky-Teukolsky identities and the Hertz map in a self-contained exposition because they will be fundamental to derive our results. In Section \ref{sec:AsymGlobalAdS}  we find the BCs on the Teukolsky variables that yields asymptotically global AdS perturbations. 
Section \ref{sec:mapKIvsTeuk} constructs the gauge invariant differential map between the Teukolsky and Kodama-Ishibashi (Regge-Wheeler$-$Zerilli) gauge invariant formalisms. Finally, in Section  \ref{sec:QNmodes}   we study the QNM spectrum / CFT timescales of the global AdS(-Schwarzschild) background.

%%%%%%%%%%%%%%%%%%%%%%%%%%%%%%%%%%%%%%%%%%%%%%%%%%%%
\section{Gravitational perturbations of the Kerr-AdS black hole \label{sec:KerrAdS}}
%%%%%%%%%%%%%%%%%%%%%%%%%%

%%%%%%%%%%%%%%%%%%%%%%%%%%%%%%%%%%%%%%%%%%%%%%%%%%%%
\subsection{Kerr-AdS black hole  \label{sec:KerrAdSbh}}
%%%%%%%%%%%%%%%%%%%%%%%%%%

The Kerr-AdS geometry was originally written by Carter in the Boyer-Lindquist coordinate system $\{ T,r,\theta,\phi\}$ \cite{Carter:1968ks}.  Here, following Chambers and Moss \cite{Chambers:1994ap}, we introduce the new time and polar coordinates $\{t,\chi\}$ related to the  Boyer-Lindquist coordinates $\{T,\theta\}$ by
\begin{equation}\label{CMtoBL}
t=\Xi\,  T \,,\qquad \chi=a \cos\theta\,,
\end{equation}
where $\Xi$ is to be defined in \eqref{metricAux}. In this coordinate system the Kerr-AdS black hole line element  reads \cite{Chambers:1994ap}
\begin{eqnarray}\label{metric}
&& ds^2=-\frac{\Delta _r}{\left(r^2+\chi ^2\right)\Xi^2}\left(dt-\frac{a^2-\chi ^2}{a}\,d\phi \right)^2+\frac{\Delta _{\chi }}{\left(r^2+\chi ^2\right)\Xi ^2}\left(dt-\frac{a^2+r^2}{a}\,d\phi\right)^2 \nonumber \\
&&\hspace{1.1cm}
+\frac{\left(r^2+\chi ^2\right)}{\Delta _r}\,dr^2+\frac{\left(r^2+\chi ^2\right)}{\Delta _{\chi}}\,d\chi^2
\end{eqnarray}
where 
\begin{equation} \label{metricAux}
 \Delta _r=\left(a^2+r^2\right)\left(1+\frac{r^2}{L^2}\right)-2M r\,,\qquad \Delta _{\chi }=\left(a^2-\chi ^2\right)\left(1-\frac{\chi ^2}{L^2}\right)\,,\qquad \Xi =1-\frac{a^2}{L^2} \,.
\end{equation}
The Chambers-Moss coordinate system $\{t,r,\chi,\phi\}$ has the nice property that the line element  treats the radial $r$ and polar $\chi$ coordinates at an almost similar footing.
One anticipates that this property will naturally extend to the radial and angular equations that describe  
gravitational perturbations in the Kerr-AdS background.
In this frame, the horizon angular velocity and temperature are given by 
\begin{equation} \label{OmegaT}
\Omega_H=\frac{a}{r_+^2+a^2}\,,\qquad 
 T_H=\frac{1}{\Xi}\left[ \frac{r_+}{2 \pi
}\left(1+\frac{r_+^2}{\ell ^2}\right)\frac{1}{r_+^2+a^2}-\frac{1}{4
\pi r_+}\left(1-\frac{r_+^2}{\ell ^2}\right)\right].
\end{equation}

The Kerr-AdS black hole obeys $R_{\mu\nu} =-3L^{-2}g_{\mu\nu}$, and asymptotically approaches global AdS space with radius of curvature $L$.  This asymptotic structure is not manifest in \eqref{metric}, one of the reasons being that the coordinate frame $\{t,r,\chi,\phi\}$  rotates at infinity with angular velocity $\Omega_{\infty}=-a/(L^2 \Xi)$.  However, if we introduce the coordinate change 
\begin{eqnarray}\label{globalAdScoordtransf}
&& T=\frac{t}{\Xi }\,, \qquad \Phi =\phi +\frac{a}{L^2}\, \frac{t}{\Xi }\,,\nonumber \\
&& R=\frac{\sqrt{L^2 \left(a^2+r^2\right)-\left(L^2+r^2\right) \chi ^2}}{L\sqrt{\Xi }},\qquad \cos\Theta=\frac{L\sqrt{\Xi } \,r \, \chi }{a\sqrt{L^2 \left(a^2+r^2\right)-\left(L^2+r^2\right) \chi ^2}}\,, 
\end{eqnarray}
we find that as $r\to\infty$ (i.e. $R\to \infty$), the Kerr-AdS geometry  \eqref{metric} reduces to
\begin{eqnarray}\label{metricAdS}
 ds^2_{AdS}=-\left(1+\frac{R^2}{L^2}\right)dT^2+\frac{dR^2}{1+\frac{R^2}{L^2}}+R^2\left(d\Theta^2+\sin^2\Theta\, d\Phi^2\right),
\end{eqnarray}
that we recognize as the line element of global AdS. In other words, the conformal boundary of the bulk spacetime is the static Einstein universe $R_t \times S^2$:  $ds^2_{bdry}=\lim_{R\to\infty} \frac{L^2}{R^2}\, ds^2_{AdS}=-dT^2+d\Theta^2+\sin^2\Theta\, d\Phi^2$. This is the boundary metric where the CFT lives in the context of the AdS$_4$/CFT$_3$ correspondence.

The ADM mass and angular momentum of the black hole are related to the mass $M$ and rotation $a$ parameters through $M_{ADM}=M/\Xi^2$
and $J_{ADM}=M a/\Xi^2$, respectively \cite{Caldarelli:1999xj,Gibbons:2004ai}. 
The horizon angular velocity and temperature that are relevant for the thermodynamic analysis  are the ones measured with respect to the non-rotating frame at infinity \cite{Caldarelli:1999xj,Gibbons:2004ai} and given in terms of \eqref{OmegaT} by $T_h= \Xi \,T_H$ and $\Omega_h=\Xi  \,\Omega _H+\frac{a}{L^2}$.
The event horizon is located at $r=r_+$ (the largest real root of $\Delta_r$), and it is a Killing horizon generated by the Killing vector $K=\partial_T+\Omega_h \partial_\Phi $. 
Further properties of the Kerr-AdS spacetime are discussed in Appendix A of~\cite{Dias:2010ma}.

%%%%%%%%%%%%%%%%%%%%%%%%%%%%%%%%%%%%%%%%%%%%%%%%%%%%
\subsection{Teukolsky master equations  \label{sec:Teukolsky}}
%%%%%%%%%%%%%%%%%%%%%%%%%%

The Kerr-AdS geometry is a Petrov type D background and therefore perturbations of this geometry can be studied using the Teukolsky formalism, which uses the Newman-Penrose (NP) framework \cite{Teukolsky:1972my,Teukolsky:1973ha,ChandrasekharBook}.  

The building blocks of this formalism are: 
\begin{itemize}

\item the NP null tetrad ${\bf e}_{a}=\{\bm{\ell},\bm{n},\bm{m},\bm{\overline{m}}\}$ (the bar demotes complex conjugation) obeying the normalization conditions $\NPl \cdot \NPn\!=\!-1,\, \NPm \cdot \NPbm\!=\!1$;

\item the NP  spin connection $\gamma_{cab}=e_b^{\phantom{b}\mu}e_c^{\phantom{c}\nu}\,\nabla_{\mu}e_{a\,\nu}$ (with $\gamma_{cab}=\!-\gamma_{acb}\,$);

\item the associated NP spin coeficients defined in terms of  $\gamma_{cab}$ as 
 \begin{fleqn}
\begin{equation}\label{NPspincoef}
\begin{alignedat}{6}
\kappa=&-\gamma_{311},\quad &\lambda=&\,\gamma_{424}, \quad&\nu=&\,\gamma_{422},\quad &\sigma=&-\gamma_{313},\quad &\alpha=&\,{\textstyle \frac{1}{2}} (\gamma_{124}-\gamma_{344}),\quad &\beta=&\,{\textstyle \frac{1}{2}} (\gamma_{433}-\gamma_{213}),\\
\mu=&\,\gamma_{423},\quad &\rho=&-\gamma_{314},\quad &\pi=&\,\gamma_{421},\quad &\tau=&-\gamma_{312},\quad  &\gamma=&\,{\textstyle \frac{1}{2}}(\gamma_{122}-\gamma_{342}),\quad &\epsilon=&\,{\textstyle \frac{1}{2}} (\gamma_{431}-\gamma_{211});
\end{alignedat}
\end{equation}
\end{fleqn}

\item the five complex Weyl scalars ($C_{abcd}$ are the Weyl tensor components in the NP null basis)
\begin{equation}\label{NPscalar}
\Psi_0= C_{1313}\,,\quad \Psi_1= C_{1213}\,,\quad \Psi_2= C_{1342}\,,\quad \Psi_3= C_{1242}\,,\quad \Psi_4 = C_{2424}\,;
\end{equation} 

\item and the NP directional derivative operators $D=\NPl^{\mu}\partial_{\mu},\,\Delta=\NPn^{\mu}\partial_{\mu},\,\delta=\NPm^{\mu}\partial_{\mu},\,\bdelta=\NPbm^{\,\mu}\partial_{\mu}\,$. 
 The complex conjugate of any complex NP quantity can be obtain through the replacement $3\leftrightarrow 4$. 
 \end{itemize}

The Kerr-AdS background is a  Petrov type D spacetime since all Weyl scalars, except $\Psi_2$, vanish: $\Psi_0=\Psi_1=\Psi_3=\Psi_4=0\,$ and $\Psi_2 = - M (r- i \chi)^{-3}$. Due to the Goldberg-Sachs theorem this further implies that $\kappa=\lambda=\nu=\sigma=0\,$. 
In addition, we might want to set $\epsilon=0$ by choosing $\NPl$ to be tangent to an affinely parametrized null geodesic $\NPl^{\mu} \nabla_{\mu} \NPl_{\nu} =0$. This was the original choice of Teukolsky and Press (when studing perturbations of the Kerr black hole) who used the the outgoing (ingoing) Kinnersly tetrad that is regular in the past (future) horizon \cite{Teukolsky:1974yv}. In the Kerr-AdS case we can work with the natural extension of Kinnersly's tetrad to AdS, and this was the choice made in \cite{Dias:2012pp}. 
However, here we choose to work with the Chambers-Moss null tetrad defined as \cite{Chambers:1994ap},
\begin{eqnarray}
 &&\NPl^{\mu}\partial_{\mu}=\frac{1}{\sqrt{2}\sqrt{r^2+\chi ^2}}\left(\Xi  \,\frac{a^2+r^2}{\sqrt{\Delta _r}} \,\partial _t+\sqrt{\Delta _r} \,\partial _r+\frac{a  \,\Xi  }{\sqrt{\Delta _r}} \,\partial _{\phi }\right),  \nonumber\\
 && \NPn^{\mu}\partial_{\mu}=\frac{1}{\sqrt{2}\sqrt{r^2+\chi ^2}}\left(\Xi   \,\frac{a^2+r^2}{\sqrt{\Delta _r}} \,\partial _t-\sqrt{\Delta _r} \,\partial _r+ \frac{a  \,\Xi }{\sqrt{\Delta _r}} \,\partial _{\phi }\right),   \nonumber\\
 &&  \NPm^{\mu}\partial_{\mu}= -\frac{i}{\sqrt{2}\sqrt{r^2+\chi ^2}}\left( \Xi  \,\frac{a^2-\chi ^2}{\sqrt{\Delta _{\chi }}} \,\partial _t+ i\sqrt{\Delta _{\chi }} \,\partial _{\chi }+\frac{a  \,\Xi }{\sqrt{\Delta _{\chi }}} \,\partial _{\phi }\right),
 \label{KerrAdS:NPtetrad}
\end{eqnarray}
which is not affinely parametrized ($\epsilon\neq 0$). The motivation for this choice is two-folded. First, the technical analysis of the angular part of the perturbation equations and solutions will be much simpler because this Chambers-Moss tetrad explores the almost equal footing treatment of the $r, \chi$ coordinates much more efficiently than Kinnersly's tetrad. Second, to complete our analysis later on we will have to discuss how the metric perturbations $h_{ab}$ (built out of the NP perturbed scalars) transform both under  infinitesimal coordinate  transformations and  infinitesimal change of basis. It turns out that if we work in the  Chambers-Moss tetrad, the results will be achieved without requiring a change of basis, while the Kinnersly's option would demand it. Again, this simplifies our exposition.

Teukolsky's treatment applies to arbitrary spin $s$ perturbations. Here, we are  interested in gravitational perturbations so we restrict our  discussion  to the  $s=\pm 2$ spins. Let us denote the unperturbed NP Weyl scalars by $\Psi_i$ and their perturbations by $\delta\Psi_i$ with $i=1,\cdots,5$. The important quantities for our discussion are the scalars $\delta\Psi_0$ and  $\delta\Psi_4$. They  are invariant both under infinitesimal coordinate transformations and under infinitesimal changes of the NP basis.
A remarkable property of the Kerr-AdS geometry is that {\it all}  information on the most {\it general} 
 \footnote{Excluding the exceptional perturbations that simply change the mass or angular momentum of the background \cite{WaldL0L1}. The Teukolsky formalism does not address these modes. See Appendix \ref{sec:completeness} for a detailed discussion.} linear perturbation of the system is encoded in these gauge invariant variables  $\delta\Psi_0$ and  $\delta\Psi_4$. That is, the perturbation of the leftover NP variables can be recovered once $\delta\Psi_0$ and  $\delta\Psi_4$ are known. The later are the solutions of the Teukolsky master equations.

For $s=2$ perturbations  the Teukolsky equation is
\begin{align}\label{TeukPosSpin}
\biggl\{&\left[D-3\epsilon+\bepsilon-4\rho-\brho\right]\left(\Delta+\mu-4\gamma\right)\nonumber\\
&\hspace{1cm}
-\left[\delta+\bpi-\balpha-3\beta-4\tau\right]\left(\bdelta+\pi-4\alpha\right)-3\Psi_2\biggr\}\,\delta\Psi_0 =4\pi\mathcal{T}_{0}\,,\quad \hbox{if}\quad  s=2,
\end{align}
while $s=-2$ perturbations are described by the Teukolsky equation
\begin{align}\label{TeukNegSpin}
\biggl\{&\left[\Delta+3\gamma-\bgamma+4\mu+\bmu\right]\left(D+4\epsilon-\rho\right)\nonumber\\
&\hspace{1cm}-\left[\bdelta-\btau+\bbeta+3\alpha+4\pi\right]\left(\delta-\tau+4\beta\right)-3\Psi_2\biggr\}\,\delta\Psi_4=4\pi\mathcal{T}_{4}\,,\quad \hbox{if}\quad  s=-2.
\end{align}
The explicit form of the source terms $\mathcal{T}_{(s^\pm)}$, that vanish in our analysis, can be found in~\cite{Teukolsky:1973ha}.

Next we introduce the separation ansatz
\begin{equation}\label{TeukAnsatz}
\left\{
\begin{array}{ll}
\delta \Psi_4=(r-i \chi )^{-2}\,e^{-i \omega  t}\,e^{i m \phi }\, R_{\omega \ell m}^{(-2)}(r)\, S_{\omega \ell m}^{(-2)}(\chi )\,, & \qquad \hbox{for}\:\:\:  s=-2 \,,\\
\delta \Psi_0=(r-i \chi )^{-2}\,e^{-i \omega  t}\,e^{i m \phi }\, R_{\omega \ell m}^{(2)}(r)\,S_{\omega \ell m}^{(2)}(\chi )\,, & \qquad \hbox{for}\:\:\: s=2 \,. 
\end{array}
\right.
\end{equation}
Also, define the radial $\{\mathcal{D}_n,\mathcal{D}^{\dagger }_n\}$ and angular $\{\mathcal{L}_n,\mathcal{L}^{\dagger}_n\}$  differential operators
\begin{eqnarray}
 &&   \mathcal{D}_n=\partial _r+i\,\frac{K_r}{\Delta _r}+n\,\frac{\Delta _r'}{\Delta _r},\qquad \mathcal{D}^{\dagger }_n=\partial_r-i\,\frac{K_r}{\Delta_r}+n\,\frac{\Delta _r'}{\Delta _r} \nonumber\\
 && \mathcal{L}_n=\partial _{\chi }+\frac{K_{\chi }}{\Delta _{\chi }}+n\,\frac{\Delta _{\chi }'}{\Delta _{\chi }}, \qquad \mathcal{L}^{\dagger}_n=\partial _{\chi }-\frac{K_{\chi }}{\Delta _{\chi }}+n\,\frac{\Delta _{\chi }'}{\Delta _{\chi }},
 \label{diffOp}
\end{eqnarray}
where the prime ${}^\prime$ represents derivative wrt the argument and
\begin{equation} \label{KrKchi}
K_r=\Xi \left[m a-\omega \left(a^2+r^2\right)\right],\qquad K_{\chi }=\Xi \left[m a -\omega \left(a^2-\chi ^2\right)\right].
\end{equation}
With the  ansatz \eqref{TeukAnsatz}, the Teukosky master equations separate into a pair of equations for the radial $R_{\omega \ell m}^{(s)}(r)$ and angular $S_{\omega \ell m}^{(s)}(\chi)$ functions, 
\begin{equation}\label{eqs:s-2}
\left\{
\begin{array}{ll}
\left(\mathcal{D}^{\dagger }_{-1}\Delta _r\mathcal{D}_{1 }+6\left(\frac{r^2}{L^2}-i\,\Xi\, \omega\,  r\right)-\lambda \right)R_{\omega \ell m}^{(-2)}(r)=0\,, & \\
\left(\mathcal{L}^{\dagger }_{-1}\Delta _{\chi }\mathcal{L}_1+6\left(\frac{\chi ^2}{L^2}+\Xi \, \omega\,  \chi \right)+\lambda \right)S_{\omega \ell m}^{(-2)}(\chi )=0\,, & \qquad \hbox{for}\:\:\:  s=-2 \,,
\end{array}
\right.
\end{equation}
\begin{equation}\label{eqs:s+2}
\hspace{-1.3cm}\left\{
\begin{array}{ll}
\left(\mathcal{D}_{-1}\Delta _r\mathcal{D}^{\dagger }_{1 }+6\left(\frac{r^2}{L^2}+i\,\Xi\, \omega\,  r \right)-\lambda\right)R_{\omega \ell m}^{(2)}(r)=0\,, & \\
\left(\mathcal{L}_{-1}\Delta _{\chi }\mathcal{L}^{\dagger }_1+6\left(\frac{\chi ^2}{L^2}-\Xi\,\omega\,\chi \right)+\lambda\right)S_{\omega \ell m}^{(2)}(\chi )=0\,, & \qquad \hbox{for}\:\:\:  s=2 \,,
\end{array}
\right.
\end{equation}
where we introduced the separation constant
\begin{equation}\label{sepConst}
\lambda \equiv \lambda_{\omega\ell m}^{(2)}=\lambda_{\omega\ell m}^{(-2)}.
\end{equation}

Some important observations are in order:

\begin{itemize}

\item First note that the radial operators obey
$\mathcal{D}^{\dagger }_n =\left(\mathcal{D}_n\right)^*$ (where ${}^*$ denotes complex conjugation) while 
the angular operators satisfy $\mathcal{L}^{\dagger}_n(\chi)=-\mathcal{L}_n(-\chi)$. 

\item Consequently, the radial  equation for $R_{\omega \ell m}^{(-2)}$ is the complex conjugate of the radial equation for $R_{\omega \ell m}^{(2)}$, but the angular solutions $S_{\omega \ell m}^{(\pm 2)}$ are instead related by the symmetry  $S_{\omega \ell m}^{(2)}(\chi)=S_{\omega \ell m}^{(-2)}(-\chi)$. The later statement implies that the separation constants are such that $\lambda_{\omega\ell m}^{(-2)}=\lambda_{\omega\ell m}^{(2)}\equiv \lambda$ with $\lambda$ being real.

\item The  eigenfunctions $S_{\omega \ell m}^{(s)}(\chi)$ are spin-weighted {\it AdS} spheroidal harmonics, with positive integer $\ell$ specifying the number of zeros, $\ell-{\rm max}\{|m|,|s|\}$ (so the smallest  $\ell$ is $\ell=|s|=2$). The associated eigenvalues $\lambda$ can be computed numerically. They are a function of $\omega,\ell,m$ and regularity imposes the constraints that $-\ell\leq m\leq \ell$ must be an integer and $\ell\geq |s|$.

\item We have the freedom to choose the normalization of the angular  eigenfunctions. A natural choice is 
\begin{equation} \label{normS}
\int_{-1}^{1} \left(S_{\omega \ell m}^{(s)}\right)^2 d\chi = 1\,. 
\end{equation}

\end{itemize}

%%%%%%%%%%%%%%%%%%%%%%%%%%%%%%%%%%%%%%%%%%%%%%%%%%%%
\subsection{Starobinsky-Teukolsky identities  \label{sec:StaroTeuk}}
%%%%%%%%%%%%%%%%%%%%%%%%%%

Suppose we solve the radial and angular equations \eqref{eqs:s+2}, \eqref{eqs:s-2} for the spin $s=\pm 2$.
These solutions for  $R_{\omega \ell m}^{(s)}$ and $S_{\omega \ell m}^{(s)}(\chi)$, when inserted in \eqref{TeukAnsatz}, are not enough to fully determine the NP gauge invariant Weyl scalars $\delta\Psi_0,\delta\Psi_4$. The reason being that the relative normalization between $\delta\Psi_0$ and $\delta\Psi_4$ remains undetermined, and thus our linear perturbation problem is yet not solved \cite{Teukolsky:1974yv,Chandra1,Chandra2}. 
Given the natural normalization \eqref{normS} chosen for the weighted spheroidal harmonics, the completion of the  solution for $\delta\Psi_0,\delta\Psi_4$ requires that we fix the relative normalization between the radial functions  $R_{\omega \ell m}^{(+2)}$ and  $R_{\omega \ell m}^{(-2)}$. This is what the Starobinsky-Teukolsky (ST) identities acomplish \cite{Staro,StaroChuri,Teukolsky:1974yv,Chandra1,Chandra2}. A detailed analysis of these identities for the Kerr black hole is available in the above original papers or in the seminal textbook of Chandrasekhar \cite{ChandrasekharBook}. Here, we present these identities for the Kerr-AdS black hole.

Act with the operator $\mathcal{D}^{\dagger }_{-1}\Delta _r \mathcal{D}^{\dagger }_0\mathcal{D}^{\dagger }_0\Delta _r\mathcal{D}^{\dagger }_1$ on the Teukolsky equation \eqref{eqs:s+2} for  $R_{\omega \ell m}^{(2)}$  and use the equation of motion for $R_{\omega \ell m}^{(-2)}$. This yields one of the radial ST identities. Similarly, to get the second, act with the operator $\mathcal{D}_{-1}\Delta _r\mathcal{D}_0\mathcal{D}_0\Delta _r\mathcal{D}_1$ on the Teukolsky equation \eqref{eqs:s-2} for  $R_{\omega \ell m}^{(-2)}$, and make use of the equation obeyed by $R_{\omega \ell m}^{(-2)}$. These radial ST identities for the Kerr-AdS background relate $R_{\omega \ell m}^{(2)}$ to $R_{\omega \ell m}^{(-2)}$,
\begin{equation}\label{StaroTeukRad}
\left\{
\begin{array}{ll}
\mathcal{D}^{\dagger }_{-1}\Delta _r \mathcal{D}^{\dagger }_0\mathcal{D}^{\dagger }_0\Delta _r\mathcal{D}^{\dagger }_1 R_{\omega \ell m}^{(2)}= \mathcal{C}_{\rm{st}} R_{\omega \ell m}^{(-2)} \,, & \\
\mathcal{D}_{-1}\Delta _r\mathcal{D}_0\mathcal{D}_0\Delta _r\mathcal{D}_1 R_{\omega \ell m}^{(-2)}=\mathcal{C}_{\rm{st}}^\ast R_{\omega \ell m}^{(2)}\,, &  
\end{array}
\right.
\end{equation}
where we have chosen the radial ST constants $\{\mathcal{C}_{\rm st},\mathcal{C}_{\rm{st}}^\ast\}$ to be related by complex conjugation. This is possible because, as noted before, the $R_{\omega \ell m}^{(\pm 2)}$ solutions are related by  complex conjugation.  

To get the angular ST identities, act with the operator $\mathcal{L}^{\dagger }_{-1}\Delta _{\mu }\mathcal{L}^{\dagger }_0\mathcal{L}^{\dagger }_0\Delta _{\mu }\mathcal{L}^{\dagger }_1$ on the Teukolsky equation \eqref{eqs:s+2} for  $S_{\omega \ell m}^{(2)}$  (and use the equation of motion for $S_{\omega \ell m}^{(-2)}$),
and act with $\mathcal{L}_{-1}\Delta _{\mu }\mathcal{L}_0\mathcal{L}_0\Delta _{\mu }\mathcal{L}_1 $ on the equation \eqref{eqs:s-2} for  $S_{\omega \ell m}^{(-2)}$ (and use the equation for $S_{\omega \ell m}^{(-2)})$. This yields the pair of ST identities,
\begin{equation}\label{StaroTeukAng}
\left\{
\begin{array}{ll}
\mathcal{L}^{\dagger }_{-1}\Delta _{\mu }\mathcal{L}^{\dagger }_0\mathcal{L}^{\dagger }_0\Delta _{\mu }\mathcal{L}^{\dagger }_1S_{\omega \ell m}^{(2)}=\mathcal{K}_{\rm st} S_{\omega \ell m}^{(-2)}\,, & \\
\mathcal{L}_{-1}\Delta _{\mu }\mathcal{L}_0\mathcal{L}_0\Delta _{\mu }\mathcal{L}_1 S_{\omega \ell m}^{(-2)}=\mathcal{K}_{\rm st} S_{\omega \ell m}^{(2)}\,. &  
\end{array}
\right.
\end{equation}
Since the equations for  $S_{\omega \ell m}^{(\pm 2)}$ are related by the symmetry $\chi \to -\chi$, the ST constant $\mathcal{K}_{\rm st}$ on the RHS of these ST identities is real. Moreover, because $S_{\omega \ell m}^{(\pm 2)}$ are both normalized to unity $-$ see \eqref{normS} $-$ the ST constant is the same in both angular ST identities.

To determine $\left| \mathcal{C}_{\rm{st}} \right|^2$ we act with the operator of the LHS of the first equation of  \eqref{StaroTeukRad} on the second equation and evaluate explicitly the resulting $8^{\rm th}$ order differential operator. A similar operation on equations \eqref{StaroTeukAng} fixes $ \mathcal{K}_{\rm{st}}^2$.
We find that
\begin{eqnarray}
 &&  \hspace{-1cm}\left| \mathcal{C}_{\rm{st}} \right|^2=\mathcal{K}_{\rm{st}}^2+144 M^2 \omega ^2\Xi ^2 \,, \label{STconstC} \\
 &&   \hspace{-1cm}\mathcal{K}_{\rm{st}}^2=\lambda^2\left(\lambda+2\right)^2+8 \lambda \Xi ^2 a \omega  \left[\left(6+5 \lambda\right)(m-a \omega )+ 12a \omega \right]+144 \Xi ^4a^2\omega ^2 (m-a \omega )^2\nonumber\\
 && \hspace{0cm}+\frac{4 a^2 }{L^2}\left[\lambda \left(\lambda+2\right)\left(\lambda-6\right)+12 \Xi ^2 (m-a \omega ) \left[2 m \lambda-a \omega \left(\lambda-6\right)\right]\right] 
 +\frac{4 a^4 \left(\lambda-6\right)^2}{L^4}\,.
 \label{STconstK}
\end{eqnarray}
This fixes completely the real constant $\mathcal{K}_{\rm{st}}$ (we choose the positive sign when taking the square root of $\mathcal{K}_{\rm{st}}^2$ to get, when $a\to 0$, the known relation between the $s=\pm 2$ spin-weighted spherical harmonics) but not  the complex constant $\mathcal{C}_{\rm{st}}$. 
However, we emphasize that to find the asymptotically global AdS boundary conditions in next  section, we do not need to know $\mathcal{C}_{\rm{st}}$, just $\mathcal{K}_{\rm{st}}$. Moreover, we do not need the explicit expression for $\mathcal{C}_{\rm st}$ to construct the map between the Kodama-Ishibashi and the $a=0$ Teukolsky formalisms of Section \ref{sec:mapKIvsTeuk}. 

Neverthless we can say a bit more about the phase of $\mathcal{C}_{\rm{st}}$. Recall that in the Kerr case,
finding  the real and imaginary parts of $\mathcal{C}_{\rm{st}}$ requires a respectful computational effort which was undertaken by Chandrasekhar \cite{Chandra2} (also reviewed in sections 82 to 95 of chapter 9 of the textbook \cite{ChandrasekharBook}).  {\it \`A priori} we would need to repeat the computations of \cite{Chandra2}, this time in the AdS background, to find the phase of  $\mathcal{C}_{\rm{st}}$ in the Kerr-AdS background (which was never done to date).  However, if we had to guess it we would take the natural assumption that $\mathcal{C}_{\rm{st}}$ is given by the solution of \eqref{STconstC} that reduces to the asymptotically flat partner of \cite{Chandra2} when $L\to \infty$, 
\begin{equation}
\mathcal{C}_{\rm st}=\mathcal{C}_1+i \,\mathcal{C}_2 \quad \hbox{with} \quad 
\mathcal{C}_1=\mathcal{K}_{\rm st}\,, \quad \mathcal{C}_2= - 12 M \omega  \Xi.
\label{STconstCf}
\end{equation}
However, we emphasize again that this expression must be read with some grain of salt and needs a derivation along the lines of \cite{Chandra2} to be fully confirmed.

Having fixed the  ST constants we have specified the relative normalization between the Teukolsky variables $\delta\Psi_0$ and  $\delta\Psi_4$. We ask the reader to see Appendix \ref{sec:completeness} for a further discussion of this issue.

%%%%%%%%%%%%%%%%%%%%%%%%%%%%%%%%%%%%%%%%%%%%%%%%%%%%
\subsection{Metric perturbations: the Hertz potentials \label{sec:Hertz}}
%%%%%%%%%%%%%%%%%%%%%%%%%%

In the previous subsections we found the solutions  of the Teukolsky master equations for the gauge invariant Weyl scalars of the Newman-Penrose formalism. We will however need to know the perturbations of the metric components, $h_{\mu\nu}=\delta g_{\mu\nu}$. These are provided by the Hertz map, $h_{\mu\nu}=h_{\mu\nu}(\psi_H)$, which reconstructs the perturbations of the metric tensor from the associated scalar Hertz potentials $\psi_H$ (in a given gauge) \cite{Cohen:1975}-\cite{Dias:2009ex}. The later are themselves closely related to the NP  Weyl scalar perturbations $\delta\Psi_0$ and  $\delta\Psi_4$.

In the Kerr-AdS background, the Hertz potentials are defined by the master equations they obey to, namely,
\begin{subequations} \label{HertzMaster}
\begin{align}
&\left[(\Delta+3\gamma-\overline{\gamma}+\overline{\mu})(D+4\epsilon+3\rho)
 -(\overline{\delta}+\overline{\beta}+3\alpha-\overline{\tau})(\delta+4\beta+3\tau)-3\Psi_{2}\right]
\psi_{H}^{(-2)}=0 \,,\\
&\left[ (D-3\epsilon+\overline{\epsilon}-\overline{\rho}) (\Delta -4\gamma-3\mu)
  -(\delta-3\beta-\overline{\alpha}+\overline{\pi}) (\overline{\delta}-4\alpha-3\pi) -3\Psi_{2}\right]
 \psi_{H}^{(2)}=0 \,.
\end{align}
\end{subequations}

Introducing the {\it ansatz} for the Hertz potential 
\begin{equation}\label{HertzAnsatz}
\psi_{H}^{(s)}= \left\{
\begin{array}{ll}
e^{-i\omega t}e^{i m\phi} (r-i \chi )^2 R_{\omega \ell m}^{(-2)}(r) S_{\omega \ell m}^{(-2)}(\chi)\,, & \qquad s=-2 \,, \\
e^{-i\omega t}e^{i m\phi} (r-i \chi )^2 R_{\omega \ell m}^{(2)}(r) S_{\omega \ell m}^{(2)}(\chi) \,, & \qquad s=2 \,,
\end{array}
\right.
\end{equation}
into \eqref{HertzMaster} (in the Kerr-AdS background), we find that $ R_{\omega \ell m}^{(s)}$ and $S_{\omega \ell m}^{(s)}$ are exactly the solutions of the radial and angular equations \eqref{eqs:s-2} and  \eqref{eqs:s+2}. This fixes the precise map between the Hertz potentials and the NP  Weyl scalar perturbations.

The Hertz map is such that the Hertz potentials $\psi_{H}^{(-2)}$ and $\psi_{H}^{(2)}$ generate the metric perturbations in two different gauges, namely the ingoing (IRG) and the outgoing (ORG) radiation gauge, defined by
\begin{equation}
\mathrm{IRG}:\;\NPl^{\mu}h_{\mu\nu}=0,\;g^{\mu\nu}h_{\mu\nu}=0\,,\qquad\mathrm{ORG}:\;\NPn^{\mu}h_{\mu\nu}=0,\;g^{\mu\nu}h_{\mu\nu}=0\,.
\end{equation}
The Hertz map is finally given by\footnote{\label{footMap}Note that \eqref{HertzORG}, whose explicit derivation can be found in an Appendix of \cite{Dias:2009ex}, corrects some typos in the map first presented in \cite{Chrzanowski:1975wv}.}
\begin{eqnarray}\label{HertzIRG}
&&h_{\mu\nu}^{\mathrm{IRG}}=\Bigl\{\NPl_{(\mu}\NPm_{\nu)}\bigl[\left(D+3\epsilon+\bepsilon-\rho+\brho\right)\left(\delta+4\beta+3\tau\right)+\left(\delta+3\beta-\balpha-\tau-\bpi\right)\left(D+4\epsilon+3\rho\right)\bigr]
\nonumber \\
&&\hspace{1cm}-\NPl_{\mu}\NPl_{\nu}\left(\delta+3\beta+\balpha-\tau\right)\left(\delta+4\beta+3\tau\right)-\NPm_{\mu}\NPm_{\nu}\left(D+3\epsilon-\bepsilon-\rho\right)\left(D+4\epsilon+3\rho\right)\Bigr\}\psi_H^{(-2)} 
\nonumber \\
&&\hspace{1cm}+\text{c.c.}\,,\\
&&h_{\mu\nu}^{\mathrm{ORG}}=\Bigl\{\NPn_{(\nu}\NPbm_{\mu)}\bigl[\left(\bdelta+\bbeta-3\alpha+\btau+\pi\right)\left(\Delta-4\gamma-3\mu\right)+\left(\Delta-3\gamma-\bgamma+\mu-\bmu\right)\left(\bdelta-4\alpha-3\pi\right)\bigr]
\nonumber \\
&&\hspace{1cm}-\NPn_{\mu}\NPn_{\nu}\left(\bdelta-\bbeta-3\alpha+\pi\right)\left(\bdelta-4\alpha-3\pi\right)-\NPbm_{\mu}\NPbm_{\nu}\left(\Delta-3\gamma+\bgamma+\mu\right)\left(\Delta-4\gamma-3\mu\right)\Bigr\}\psi_H^{(2)} \nonumber \\
&&\hspace{1cm}+\text{c.c.}\,.
\label{HertzORG}
\end{eqnarray}
We have explicitly checked that \eqref{HertzIRG} and \eqref{HertzORG} satisfy the linearized Einstein equation (see also footnote \ref{footMap}).

It is important to emphasize that the Hertz map provides the {\it most general} metric perturbation with $\ell\geq 2$ of the Kerr-AdS black hole \cite{Teukolsky:1973ha,Chandra1,Chandra2,Wald}.
We defer a detailed discussion of this  observation to Appendix \ref{sec:completeness}. 

%%%%%%%%%%%%%%%%%%%%%%%%%%%%%%%%%%%%%%%%%%%%%%%%%%%%
\section{Boundary conditions for global AdS perturbations  of Kerr-AdS \label{sec:AsymGlobalAdS}}
%%%%%%%%%%%%%%%%%%%%%%%%%%

We start this section with a brief recap of the Teukolsky system which emphasizes some of its properties that are essential to discuss the asymptotic boundary conditions.

The gravitational Teukolsky equations are described by a set of two families of equations, one for spin $s=2$ and the other for $s=-2$. In Petrov type D backgrounds, these two families encode the same information, once we use the Starobinsky-Teukolsky identities that fix the relative normalization between both spin-weighted spheroidal harmonics $S_{\omega \ell m}^{(s)}$ and between both  radial functions $R_{\omega \ell m}^{(s)}$. Indeed, modulo the ST relative normalization, the two radial functions are simply the complex conjugate of each other, and the two angular functions are related by $S_{\omega \ell m}^{(2)}(\chi)=S_{\omega \ell m}^{(-2)}(-\chi)$. This is a consequence of the fact that the Teukolsky operator acting on  $\delta\Psi_0$ is the adjoint of the one acting on $\delta\Psi_4$.

The upshot of these observations, with relevance for practical applications, is that the Teukosky system in Petrov type D geometries is such that we just need to analyze the $s=2$ sector (for example) to find all the information, except BCs, on the gravitational perturbations (excluding modes that just shift the mass and angular momentum).  In other words, given $R_{\omega \ell m}^{(2)}$, $S_{\omega \ell m}^{(2)}$ and the ST constants we can reconstruct all the $s=-2$ Teukolsky quantities.
 
Were we discussing perturbations of the asymptotically flat Kerr black hole and this section on the boundary conditions would end with the following single last observation. Being a second order differential system, the gravitational field has two independent asymptotic solutions, namely, the ingoing and outgoing traveling modes. Since we are not interested in scattering experiments, the BC (that preserves asymptotic flatness) would be fixed by selecting the purely outgoing BC. For practical purposes, we would definitely just need to study the $s=2$ Teukolsky system of equations. 
 
The situation is far less trivial when we look into perturbations of Kerr-AdS. This time the second order differential system has two independent asymptotic solutions that are power laws of the radial variable. 
The BC to be chosen selects the relative normalization between these two solutions. What is the criterion to make this choice? This will be made precise in the next subsection. Before such a formal analysis we can however describe it at the heuristic level. Basically we want the perturbed background to preserve the asymptotic global AdS character of the Kerr-AdS background. Global AdS asymptotic structure  means that the system behaves as a confining box were the only allowed perturbations are those described by standing waves. Standing waves on the other hand can be decomposed as a fine-tuned sum of IRG and ORG modes such that we have a node at the asymptotic AdS wall. With this brief argument we conclude that to find the asymptotic global AdS BC we necessarily need to use the information on {\it both} the IRG and ORG Teukolsky metric perturbations, i.e. the BC discussion will require using information on both spins. Once we find it, it is still true that the spin $s=2$ sector of the Teukolsky system encodes the same information as the $s=-2$ one, and we will be able to study the properties of perturbations in Kerr-AdS using only the $s=2$ sector (say).  (Note that an infinitesimal rotation of the tetrad basis should allow to derive our results using only the ORG, say).

So we take the most general gravitational perturbation of the Kerr-AdS black hole to be given by the sum of the ingoing and outgoing radiation gauge contributions as written in \eqref{HertzIRG} and \eqref{HertzORG}. (By diffeomorphism invariance, this solution can be written in any other gauge through a gauge transformation). 
The physically relevant perturbations are those that are regular at the horizon and asymptotically global AdS. In this section we find one of our most fundamental results, namely the BCs we need to impose on our perturbations.

%%%%%%%%%%%%%%%%%%%%%%%%%%%%%%%%%%%%%%%%%%%%%%%%%%%%
\subsection{Definition of asymptotically global AdS perturbations  \label{sec:HT}}
%%%%%%%%%%%%%%%%%%%%%%%%

When considering linear perturbations of a background we have in mind two key properties: the perturbations should keep the spacetime regular and they should be as generic as possible, but without being so violent that they would destroy the asymptotic structure of the background. To make this statement quantitative, in the familiar case of an asymptotically Minkowski background, the appropriate boundary condition follows from the requirement that the perturbations preserve asymptotically the Poincar\'e group of the Minkowski spacetime \cite{Geroch}.   

For the AdS case, Boucher, Gibbons, and Horowitz \cite{Boucher:1983cv} and Henneaux and Teiltelboim \cite{Henneaux:1985tv}  have defined precisely what are the asymptotic BC we should impose to get perturbations that approach at large spacelike distances the global AdS spacetime. The main guideline is that  perturbations in a global AdS background  must preserve asymptotically  the global AdS symmetry group $O (3, 2)$, much like perturbations in a flat background must preserve asymptotically  the Poincar\'e group of the Minkowski spacetime.
More concretely, asymptotically global AdS spacetimes are defined by BCs on the gravitational field which obey the following three requirements \cite{Henneaux:1985tv}:

\noindent $\hspace{0.2cm}$ 
 (1) they should contain the asymptotic decay of the Kerr-AdS metric;

\noindent $\hspace{0.2cm}$ 
 (2) they should be invariant under the global AdS symmetry group $O (3, 2)$;

\noindent $\hspace{0.2cm}$
 (3) they should make finite the surface integral charges associated with the  $O(3, 2)$ generators.

If we work in the coordinate system $\{ T,R,\Theta,\Phi\}$, where the line element of global AdS is given by \eqref{metricAdS}, the metric perturbations that obey the above BCs behave asymptotically as~\cite{Henneaux:1985tv}:
\begin{subequations} \label{BCsHT}
\begin{align}
& h_{T\mu}=\frac{1}{R}\,F_{T\mu}(T,\Theta,\Phi)+\mathcal{O}\left( R^{-2}\right)\,, \qquad \hbox{for}\quad \mu=T,\Theta,\Phi\,,\\
& h_{TR}=\frac{1}{R^4}\,F_{TR}(T,\Theta,\Phi)+\mathcal{O}\left( R^{-5}\right)\,,\\
& h_{RR}=\frac{1}{R^5}\,F_{RR}(T,\Theta,\Phi)+\mathcal{O}\left( R^{-6}\right)\,,\\
& h_{R\mu}=\frac{1}{R^4}\,F_{R\mu}(T,\Theta,\Phi)+\mathcal{O}\left( R^{-5}\right)\,, \qquad \hbox{for}\quad \mu=\Theta,\Phi\,,\\
& h_{\mu\nu}=\frac{1}{R}\,F_{\mu\nu}(T,\Theta,\Phi)+\mathcal{O}\left( R^{-2}\right)\,, \qquad \hbox{for}\quad \mu,\nu=\Theta,\Phi\,,
\end{align}
\end{subequations}
where $F_{\mu\nu}(T,\Theta,\Phi)$ are functions of $\{T,\Theta,\Phi\}$ only.

These BCs are defined with respect to a particular coordinate system. Consider a generic infinitesimal coordinate transformation $x^\mu\to x^\mu+\xi^\mu$, where $\xi$ is an arbitrary gauge vector field. 
Under this gauge transformation the metric perturbation transforms according to
\begin{equation}\label{gaugeTransf}
h_{\mu\nu} \to  h_{\mu\nu} - 2\nabla_{(\mu}\xi _{\nu)}\,,
\end{equation}
which we can use to translate the BCs \eqref{BCsHT} in the $\{ T,R,\Theta,\Phi\}$ frame into any other coordinate system, so long as $\xi$ decays sufficiently fast at infinity.

%%%%%%%%%%%%%%%%%%%%%%%%%%%%%%%%%%%%%%%%%%%%%%%%%%%%
\subsection{Boundary conditions for asymptotically global AdS perturbations  \label{sec:BCinf}}
%%%%%%%%%%%%%%%%%%%%%%%%%%
Modulo gauge transformations, the most general perturbation of linearized Einstein equations in the Kerr-AdS background can be written as  
\begin{equation}\label{generalSol}
h_{\mu\nu}=h_{\mu\nu}^{\rm IRG}+h_{\mu\nu}^{\rm ORG},
\end{equation}
where $h_{\mu\nu}^{\rm IRG}$ and  $h_{\mu\nu}^{\rm ORG}$  are determined by the Hertz maps \eqref{HertzIRG} and \eqref{HertzORG}, with the Hertz potentials $\psi_{H}^{(\pm 2)}$ defined in \eqref{HertzAnsatz} and the associated Teukolsky functions obeying the equations of motion   \eqref{eqs:s+2} and  \eqref{eqs:s-2}. Note that the relative normalization between these two contributions is fixed by the  Starobinsky-Teukolsky treatment.

Solving the radial Teukolsky equations  \eqref{eqs:s+2} and  \eqref{eqs:s-2} at infinity, using a standard Frobenius analysis, we find that the two independent asymptotic decays for $ R_{\omega \ell m}^{(\pm 2)}$ are
\begin{eqnarray}\label{FrobR}
&&  R_{\omega \ell m}^{(2)}{\bigl |}_{r\to \infty}\sim  A_{+}^{(2)}\,\frac{L}{r}+ A_{-}^{(2)}\,\frac{L^2}{r^2}+\mathcal{O}\left( \frac{L^3}{r^3}\right)\,,\nonumber \\
&&  R_{\omega \ell m}^{(- 2)}{\bigl |}_{r\to \infty}\sim B_{+}^{(- 2)}\,\frac{L}{r}+ B_{-}^{(-2)}\,\frac{L^2}{r^2}+\mathcal{O}\left(  \frac{L^3}{r^3} \right)\,,
\end{eqnarray}
where  the amplitudes $A_{\pm}^{(s)}\equiv \{A_{\pm}^{(2)},B_{\pm}^{(-2)}\}$ are, at this point, independent arbitrary constants. Our task is to find the BC  we have to impose  in order to get a perturbation $h_{\mu\nu}$ that is asymptotically global AdS. That is, we must find  the constraints,  $A_{-}^{(s)}= A_{-}^{(s)}\left( A_{+}^{(s)} \right)$,  that these amplitudes have to obey to get the Henneaux-Teiltelboim decay \eqref{BCsHT}.  We will find that the most tempting condition, where we set to zero the leading order term in the expansion, $A_{+}^{(s)}=0$, is too naive and does not do the job.
Note that it follows from \eqref{globalAdScoordtransf} that for large $R$, or $r$, one has $R\sim r \left[ (L^2-\chi^2)/L^2\Xi \right]^{1/2}$ and $\cos\Theta\sim\chi \left[ L^2\Xi /(L^2-\chi^2)\right]^{1/2}$. Therefore, to get the asymptotically global AdS decay of $h_{\mu\nu}$ in the $\{t,r,\chi,\phi\}$ coordinate system we can simply replace $\{T,R,\Theta,\Phi\}\to \{t,r,\chi,\phi\}$ in   \eqref{BCsHT}.

 In $h_{\mu\nu}^{\rm IRG}$, we can express $S_{\omega \ell m}^{(-2)}(\chi)$ as a function of $S_{\omega \ell m}^{(2)}(\chi)$ and its derivative using the first Starobinsky-Teukolsky (ST) identity in \eqref{StaroTeukAng} and the angular equation of motion \eqref{eqs:s+2}. This eliminates $S_{\omega \ell m}^{(-2)}$ from \eqref{generalSol}. At this stage, we could also immediately replace the ST constant $\mathcal{K}_{\rm st}$ by its expression \eqref{STconstK}. Instead, we choose to keep it unspecified until a later stage in our computation. 

The explicit expression of $h_{\mu\nu}^{\rm IRG}+h_{\mu\nu}^{\rm ORG}$ when we introduce \eqref{FrobR} into  \eqref{HertzIRG} and \eqref{HertzORG} contains order $r^2$ terms but no other higher power of $r$. Our first task is to use all the gauge freedom \eqref{gaugeTransf} to eliminate, if possible, these  $\mathcal{O}(r^2)$ terms  and all lower power law terms that are absent in the asymptotically global AdS decay  \eqref{BCsHT}. The gauge parameter compatible with the background isometries is $\xi=e^{-i \omega t}e^{i m \phi} \xi_\mu(r,\chi) dx^\mu$.
A simple inspection of $\nabla_{(\mu}\xi _{\nu)}$ concludes that the most general components of the gauge vector field, that can contribute up to $\mathcal{O}(r^2)$ terms, can be written as the power law expansion in $r$:
\begin{equation}
 \xi_t=\sum_{j=0}^{n} \xi_t^{(j)}(\chi) r^{2-j}\,,\quad  \xi_r=\sum_{j=0}^{n} \xi_r^{(j)}(\chi) r^{-(1+j)}\,,\quad \xi_\chi=\sum_{j=0}^{n} \xi_\chi^{(j)}(\chi) r^{2-j}\,,\quad \xi_\phi=\sum_{j=0}^{n} \xi_\phi^{(j)}(\chi) r^{2-j}. 
\end{equation}
Inserting this expansion and  \eqref{FrobR}  into \eqref{generalSol}, we find that there is a judicious choice of the functions  $\xi_\mu^{(i)}(\chi)$ such that we can eliminate most of the radial power law terms that are absent in the several metric components of \eqref{BCsHT} (the expressions are long and not illuminating).
More concretely, we are able to gauge away all desired terms but the $\mathcal{O}(r^2)$ contribution in the components $h_{\chi\chi}$, $h_{\chi\phi}$ and $h_{\phi\phi}$. 

At this point, having used all the available diffeomorphism \eqref{gaugeTransf}, we find ourselves at a key stage of the analysis. To eliminate the undesired leftover $\mathcal{O}(r^2)$ contributions  we will have to fix the BCs that the amplitudes introduced in \eqref{FrobR} have to obey to guarantee that the perturbation is asymptotically global AdS. There are two conditions that eliminate simultaneously the $\mathcal{O}(r^2)$ terms in $h_{\chi\chi}$, $h_{\chi\phi}$ and $h_{\phi\phi}$. One is the coefficient of a term proportional to $S_{\omega \ell m}^{(2)}(\chi)$, and the other is proportional to $\partial_\chi S_{\omega \ell m}^{(2)}(\chi)$.\footnote{Note that in this computation we use the angular equation of motion \eqref{eqs:s+2}  to get rid of second and higher derivatives of $S_{\omega \ell m}^{(2)}$.} Clearly, these two contributions have to vanish independently. We can use them to express, for example,  the amplitude $B_{+}^{(- 2)}$ and the ST  constant $\mathcal{K}_{st}$ in terms of the other amplitudes $B_{-}^{(-2)}, A_{\pm}^{(2)}$, perturbation parameters $ \omega, m, \lambda$ and  the rotation background parameter $a$ (the mass parameter $M$ is absent in these expressions):  
 \begin{eqnarray}
&& 0= W B_+^{(-2)}-B_-^{(-2)}L {\biggl[}-4\,a\,m\,L\,\Xi  \left(2 \,i \,A_-^{(2)}+5 \,L\,\Xi\,\omega  A_+^{(2)}\right)+2 a^2\left(-6+\lambda +6L^2 \Xi ^2 \omega ^2\right)A_+^{(2)}\nonumber \\
&& \hspace{0.7cm} + \lambda \,L ^2\left(2+\lambda -4L ^2 \Xi ^2 \omega ^2\right)A_+^{(2)}-8L ^3 \Xi  \,\omega \left(i A_-^{(2)}+L  \,\Xi \, \omega  A_+^{(2)}\right) \left(1+\frac{1}{2}\lambda -L ^2 \Xi ^2 \omega ^2\right)   {\biggr]},  \nonumber \\
&&  \label{condA} \\
&& \mathcal{K}_{\rm{st}}= -\frac{B_-^{(-2)}}{L ^3 W} {\biggl\{}   \lambda ^2 (2+\lambda )^2L ^6+8 a\, \lambda  (6+5 \lambda ) m \,L^6 \Xi ^2 \omega -144 a^3 m \,L ^4 \Xi ^2 \omega  \left(-2+\lambda +2L ^2 \Xi ^2 \omega ^2\right)
 \nonumber \\
 &&   \hspace{2.6cm} +4 a^2L ^4  {\biggl [} \lambda  \left[-12+(-4+\lambda ) \lambda +24 m^2\right]+2 \left[(6-5 \lambda ) \lambda +18 m^2\right]L ^2 \Xi ^2 \omega ^2 {\biggr ]} 
 \nonumber \\
 &&   \hspace{2.6cm} +4 a^4L ^2  {\biggl [} 36-12 \lambda +\lambda ^2-48 \lambda  m^2+12 \left[\lambda -6 \left(1+m^2\right)\right]L ^2 \Xi ^2 \omega ^2+36L ^4 \Xi ^4 \omega ^4{\biggr ]} 
  \nonumber \\
 &&  \hspace{2.6cm} +48 a^6 m^2 \left(2 \lambda +3L ^2 \Xi ^2 \omega ^2\right) 
 {\biggr \}},  \label{condK} 
  \end{eqnarray}
where we have defined
\begin{eqnarray}
&& W \equiv 
L ^3\left[\lambda  (2+\lambda ) A_-^{(2)}-4 (1+\lambda )L  \,\Xi \, \omega  \left(i \lambda A_+^{(2)}+2L  \,\Xi \, \omega  A_-^{(2)}\right)+4 i (2+3 \lambda )L ^3 \Xi ^3 \omega ^3 A_+^{(2)}\right]
 \nonumber \\
 &&  \hspace{1cm}  
  +8L ^7 \Xi ^4 \omega ^4 \left(A_-^{(2)}-i\,L \, \Xi \, \omega  A_+^{(2)}\right)-4 \,a\, m \,L ^2 \Xi \left[3\, i \,\lambda  A_+^{(2)}+L \, \Xi \, \omega  \left(5 A_-^{(2)}-8\, i\,L \, \Xi \, \omega  A_+^{(2)}\right)\right] \nonumber \\
 &&  \hspace{1cm}  
+2 a^2L  \left[2\left(2\, i\,L \, \Xi \, \omega  A_+^{(2)}-A_-^{(2)}\right) \left(3+\lambda -3\,L^2 \Xi ^2 \omega ^2\right)+3\lambda  A_-^{(2)}\right].
 \end{eqnarray}

At this stage we finally introduce the explicit expression for the angular Starobinski-Teukolsky constant, namely,  $\mathcal{K}_{\rm{st}}$ is given by the positive square root of \eqref{STconstK}.
In addition, we also use the property that the radial  $R_{\omega \ell m}^{(\pm 2)}$ solutions are the complex conjugate of each other. 
In these conditions we find that conditions \eqref{condA}-\eqref{condK} are obeyed if and only if
 \begin{equation}\label{TheBCs}
 A_-^{(2)}=-i \, \eta A_+^{(2)} \,,\qquad  B_-^{(-2)}=i \,\eta B_+^{(-2)} \,,
\end{equation}
with two possible solutions for $\eta$, that we call $\eta_{\rm \bf s}$ and $\eta_{\rm \bf v}$, for reasons that will become clear in the next section. These define the BCs we look for.
To sum up, the two possible BCs on the Teukolsky amplitudes, defined in  \eqref{FrobR},  that yield an asymptotically global AdS perturbation, take the form  \eqref{TheBCs} with
 \begin{eqnarray}
&& 1)   \quad  \eta=\eta_{\rm \bf s}= \frac{\Lambda_0-\sqrt{\Lambda_1}}{\Lambda_2}\,,\qquad \hbox{or} \label{TheBCsS} \\
 && 2) \quad \eta=\eta_{\rm \bf v}= \frac{\Lambda_0+\sqrt{\Lambda_1}}{\Lambda_2}\,,\label{TheBCsV}
 \end{eqnarray}
where we have introduced
\begin{eqnarray}  \label{TheBCaux}
&& \Lambda_0 \equiv 2 a^2 (\lambda -6)-8 (\lambda +1) L^4 \omega ^2\Xi ^2+8 L^6 \omega ^4\Xi ^4+L^2 \left[\lambda  (\lambda +2)-4\text{  }\Xi ^2a \omega  \left[5 (m-a \omega )+2 a \omega \right]\right], \nonumber \\
 &&  \Lambda_1 \equiv  4 a^4 (\lambda -6)^2+L^4\lambda ^2 (\lambda +2)^2+48 (\lambda +6)a^3 \Xi ^2L^2 \omega (m-a \omega )
  +8\lambda  (5 \lambda +6)(m-a \omega ) L^4\Xi ^2 a \omega 
   \nonumber\\
 && \hspace{0.9cm} +4 a^2 L^2 {\biggl [} \lambda  \left[-12+(\lambda -4) \lambda +24 (m-a \omega )^2 \Xi ^2\right] 
 +12  \Xi ^2 L^2 \omega ^2 \left[2 \lambda +3 (m-a \omega )^2 \Xi ^2\right] {\biggr ]}, \nonumber\\
 &&\Lambda_2 \equiv 4 L \Xi \left[2 a m+L^2 \omega  \left(2+\lambda -2 L^2 \omega ^2\Xi ^2\right)\right].
 \end{eqnarray}
 Note that the BCs do not depend on the mass parameter $M$ of the background black hole (neither does the ST constant $\mathcal{K}_{\rm{st}}$).
 
The metric of the Kerr-AdS black hole asymptotically approaches that of global AdS. The boundary conditions \eqref{TheBCs}-\eqref{TheBCsS}  are the most fundamental result of our study: perturbations obeying these BCs are the ones that preserve the asymptotically global AdS behavior of the background.
These are also natural BCs in the context of the AdS/CFT correspondence: they allow a non-zero expectation value for the CFT stress-energy tensor while keeping fixed the boundary metric. The reader interested in different BCs that allow, e.g. for a dynamical boundary metric, can start from the respective asymptotic metric decay that replaces \eqref{BCsHT} and work out the above procedure to get the associate BCs on the Teukoslsky variables.

%%%%%%%%%%%%%%%%%%%%%%%%%%%%%%%%%%%%%%%%%%%%%%%%%%%%
\subsection{Horizon boundary conditions \label{sec:BChor}}
%%%%%%%%%%%%%%%%%%%%%%%%%%

At the horizon, the BCs must be such that only ingoing modes are allowed.

A Frobenious analysis at the horizon gives the two independent solutions,
\begin{equation}\label{bcH0}
R_{\omega \ell m}^{(s)}\sim  A_{\rm{in}}\left(r-r_+\right){}^{-\frac{s}{2}-i\frac{ \omega -m \Omega _H}{4\pi  T_H}}+A_{\rm{out}}\left(r-r_+\right){}^{\frac{s}{2}+i\frac{ \omega -m \Omega _H}{4\pi  T_H}}
+\mathcal{O}\left( (r-r_+)^2 \right), 
\end{equation}
where $A_{\rm{in}},A_{\rm{out}}$ are arbitrary amplitudes and $\Omega_H,T_H$ are the angular velocity and temperature defined in \eqref{OmegaT}.
The BC is determined by the requirement that the solution is regular in ingoing Eddington-Finkelstein coordinates (appropriate to extend the analysis through the horizon) and demanding regularity of the Teukolsky variable in this coordinate system. This requires that we set $A_{\rm{out}}=0$ in \eqref{bcH0}:
\begin{equation}\label{bcH}
R_{\omega \ell m}^{(s)}{\biggl |}_{r\to r_+}=A_{\rm{in}}\left(r-r_+\right)^{-\frac{s}{2}-i\frac{ \omega -m \Omega _H}{4\pi  T_H}} +\mathcal{O}\left( (r-r_+)^2 \right).
\end{equation}

%%%%%%%%%%%%%%%%%%%%%%%%%%%%%%%%%%%%%%%%%%%%%%%%%%%%
\section{Map between Teukolsky and Kodama-Ishibashi formalisms ($a=0$)\label{sec:mapKIvsTeuk}}
%%%%%%%%%%%%%%%%%%%%%%%%%%
  
In the previous section we found the boundary conditions we need to impose on the solution of the Teukolsky master equation to get gravitational perturbations of the Kerr-AdS black hole  that preserve the asymptotically  global AdS behavior of the background. The Kerr-AdS family includes the global AdS-Schwarzchild black hole and the global AdS geometry as special elements when we set, respectively,  $a=0$ and $a=0=M$. Thus, our BCs also apply to perturbations of these static backgrounds. 

On the other hand, perturbations of the global AdS(-Schwarzchild)   backgrounds were already studied in great detail in the literature using the Kodama-Ishibashi (KI) gauge invariant formalism. In four dimensions, the KI formalism reduces exactly to the analysis firstly done by Zerilli and Regge and Wheeler (in the $L\to\infty$ case). Indeed, the KI vector master equation is the Regge-Wheeler master equation for odd (also called axial) perturbations \cite{Regge:1957td}, and the KI scalar master equation is the Zerilli master equation for even (also called polar) perturbations \cite{Zerilli:1970se} .

Clearly, there must be a one-to-one map between the Kodama-Ishibashi and the Teukolsky  formalisms (when $a=0$). This map was never worked out so we take the opportunity to find it\footnote{For an earlier discussion of the relation between the Teukolsky and the Regge-Wheeler$-$Zerilli  variables in the asymptotically flat case see \cite{Lousto}.}. Actually, this task will reveal to be quite fruitful since we will find some remarkably simple connections.  

Recall the main difference between the Kodama-Ishibashi and the Teukolsky  formalisms. The former  
is well suited only for backgrounds that are locally the product of a 2-dimensional orbit spacetime (parametrized by the time and radial coordinates) and a base space $\mathcal{K}$. If this is the case, we can do a harmonic decomposition of the perturbations $h_{ab}$ according to how they transform under coordinate transformations on $\mathcal{K}$. This is certainly the case of  the global AdS(-Schwarzchild) backgrounds where the base space is a sphere,  $\mathcal{K}=S^{2}$, and we can introduce a {\it spherical} harmonic decomposition of gravitational perturbations. (Unfortunately, the Kerr-AdS geometry cannot be written as a local product of two such spaces and the KI formalism does not apply to it).
On the other hand, the Teukolsky  formalism uses a harmonic decomposition with respect to the {\it spin-weighted spherical} harmonics $S_{\omega\ell m}^{(s)}$. 

These two harmonic decompositions are distinct and responsible for the differences between the Teukolsky and KI formalisms. We can however write uniquely the scalar/vector KI harmonics in terms of  standard scalar spherical harmonics (associated Legendre polynomials), and there is another unique differential map that generates the spin-weighted spherical harmonics also from  standard spherical harmonics. This provides the necessary bridge between the two formalisms that leads to their unique map. To appreciate this  we will discuss these harmonics in detail. Also, to make the KI discussion self-contained, we will briefly review the KI formalism in the next subsection, before constructing the desired map in subsection \ref{sec:KIvsTeuk}.  

So far we have not discussed the role of the boundary conditions (BCs) in this map. We found a set of two distinct BCs for the Teukolsky solution. Quite interestingly, we will see that Teukolsky perturbations with BC \eqref{TheBCsS} maps to the KI scalar modes while  Teukolsky perturbations with BC \eqref{TheBCsV} generates the  KI vector modes.

%%%%%%%%%%%%%%%%%%%%%%%%%%%%%%%%%%%%%%%%%%%%%%%%%%%%
\subsection{Kodama-Ishibashi gauge invariant formalism \label{sec:KI}}
%%%%%%%%%%%%%%%%%%%%%%%%%%

In the Kodama-Ishibashi (KI) formalism  \cite{Kodama:2003jz}, the  most general perturbation of  the global AdS(-Schwarzchild)  geometries is decomposed  into a superposition of two classes of modes:  scalar and vector. Scalar and vector modes are expanded in terms of the scalar $\scalar(x,\phi)$  and vector $\vector_j(x,\phi)$  harmonics that we review next ($x=\cos\theta$; see \eqref{CMtoBL}).

Scalar perturbations are given by   \cite{Kodama:2003jz}
\begin{eqnarray}\label{pert:scalar}
  h^{\rm \bf s}_{ab}= f_{ab} \scalar, \qquad  
  h^{\rm \bf s}_{ai}= rf_a \scalar_i  , \qquad 
  h^{\rm \bf s}_{ij}= 2r^2 \left( H_L\gamma_{ij}\scalar + H_T \scalar_{ij} \right),
\end{eqnarray}
where $(a,b)$ are components in the orbit spacetime parametrized by $\{t,r\}$, $(i,j)$ are legs on the sphere,  $\{f_{ab},f_a,H_T,H_L\}$ are functions of  $(t,r)$, and $\scalar$ is the KI scalar harmonic, 
$ \scalar_i = -\lambda_{\rm \bf s}^{-1/2} D_i \scalar$,   $S_{ij} = \lambda_{\rm \bf s}^{-1} D_i D_j \scalar + \frac{1}{2}\gamma_{ij} \scalar$,
$\gamma_{jk}$ is the unit radius metric on $S^2$  and $D_j$ is the associated covariant derivative.
Assuming the ansatz $S(x,\phi)=e^{i m \phi}Y_{\ell_{\rm \bf s}}^{m}(x)$, the scalar harmonic equation $\left (\triangle_{S^2} + \lambda_{\rm \bf s} \right)\scalar=0$  ($\triangle_{S^2} = \gamma^{jk}  D_j D_k$) reduces to
\begin{equation}\label{KI:scalarHeq}
\partial _x\left[\left(1-x^2\right)\partial _x Y_{\ell_{\rm \bf s}}^{m}(x)\right]+\left(\lambda_{\rm \bf s} -\frac{m^2}{1-x^2}\right)Y_{\ell_{\rm \bf s}}^{m}(x)=0\,.   
\end{equation}
Its regular solutions, with normalization $\int _0^{2\pi }d\phi\int _{-1}^1 dx\,\left | \scalar \right|^2 =1$, are
\begin{eqnarray}\label{KI:scalarH}
&& \scalar(x,\phi)=\sqrt{\frac{2\ell_{\rm \bf s} +1}{4\pi }\frac{(\ell_{\rm \bf s} -m)!}{(\ell_{\rm \bf s} +m)!}}\, P_{\ell_{\rm \bf s}}^m(x)\,e^{i m \phi}\equiv  Y_{\ell_{\rm \bf s}}^{m}(x,\phi)\,, \\ 
&& \hbox{with} \quad \lambda_{\rm \bf s}=\ell_{\rm \bf s}\left(\ell_{\rm \bf s}+1\right),\qquad  \ell_{\rm \bf s}=0,1,2,\cdots,\quad |m|\leq  \ell_{\rm \bf s}.\nonumber
\end{eqnarray}
where $P_{\ell}^m(x)$ is the associated Legendre polynomial. Hence, the KI scalar harmonic $ \scalar(x,\phi)$ is the standard scalar spherical harmonic $Y_{\ell_{\rm \bf s}}^{m}(x,\phi)$.

On the other hand, the KI  vector perturbations are given by  \cite{Kodama:2003jz}
\begin{equation}\label{pert:vector}
   h^{\rm \bf v}_{ab}=0 \,, \qquad 
  h^{\rm \bf v}_{ai}=r h_a \vector_i ,\qquad 
  h^{\rm \bf v}_{ij} = -2 \lambda_{\rm \bf v}^{-1} r^2 h_T  D_{(i}\vector_{j)} \,, 
\end{equation} 
where $\{h_a, h_T\}$ are functions of $(t,r)$ and the KI  vector harmonics $\vector_j$ are the  solutions of 
\begin{equation}\label{KI:vectorHeq}
\left (\triangle_{S^2} + \lambda_{\rm \bf v} \right)\vector_j=0 \,, \qquad D_j \vector^j=0 \quad 
   \hbox{(Transverse condition)}.
\end{equation} 
The regular vector harmonics can be written in terms of the spherical harmonic $Y_{\ell_{\rm \bf v}}^{m}(x,\phi)$ as
\begin{eqnarray}\label{KI:vectorH}
&& \vector_j\,dx^j=-\frac{i \,m}{1-x^2}Y_{\ell_{\rm \bf v}}^{m}(x,\phi )\,dx+\left(1-x^2\right)\partial_x Y_{\ell_{\rm \bf v}}^{m}(x,\phi )\,d\phi\,,  \\
&& \hbox{with}    \qquad \lambda_{\rm \bf v}=\ell_{\rm \bf v}\left(\ell_{\rm \bf v}+1\right)-1,\qquad  \ell_{\rm \bf v}=1,2,\cdots,\quad |m|\leq  \ell_{\rm \bf v}.\nonumber
\end{eqnarray}

With this harmonic decomposition, the linearized Einstein equation reduces to a set of two decoupled gauge invariant KI master equations for the KI master fields $ \Phi^{(j)}_{\ell_j}$ which can be written in a compact form as,\begin{equation}
 \left(  \Box_2 - \frac{U_j}{f} \right) \Phi^{(j)}_{\ell_j}(t,r)=0 \,, \qquad  \hbox{where} \quad f=1+\frac{r^2}{L^2}-\frac{2M}{r}\,, \quad \hbox{and}\quad j=\{ {\rm \bf s},{\rm \bf v}\},
\label{KImaster}
\end{equation}
for scalar (${\rm \bf s}$), and vector (${\rm \bf v}$) perturbations. Here, $\Box_2$ is the d'Alembertian operator in the 2-dimensional orbit spacetime, and the expression for the potentials $\{U_{\rm \bf s},U_{\rm \bf v}\}$ can be found in equations (3.2)-(3.8) and  (5.15)  of  KI  \cite{Kodama:2003jz}, respectively.
They depend on the properties of the background, namely on the mass parameter $M$ and cosmological length $L$, and on the eigenvalues $\lambda_{\rm \bf s}, \lambda_{\rm \bf v}$ of the associated (regular) vector, and scalar harmonics defined above.
Since the background is  time-translation invariant,  the fields can be further Fourier decomposed in time as $ \Phi^{(j)}_{\ell_j}(t,r)= e^{-i\omega_j t} \Phi^{(j)}_{\omega_j \ell_j} (r)$.

We will need to express  KI master fields in terms of the metric functions. Going through  \cite{Kodama:2003jz}, for $\omega\neq 0$, one finds that the gauge invariant KI scalar master field is given by
\begin{eqnarray}\label{KI:scalarPhi}
&&\hspace{-0.3cm}\Phi^{\rm (s)}_{\omega \ell_{\rm \bf s}}= \frac{ - \omega \,r \left(f_{tt}+4f H_L\right)+f^2 \left(2 i f_{tr}+ \omega \, r f_{rr}\right)}{ \omega \, f \left(\lambda_{\rm \bf s}-2 f+r f' \right)}+\frac{2 i \left[r f^2 f_t'+\left(r^2 \omega ^2+f^2-r f f'\right)f_t\right]}{\sqrt{\lambda_{\rm \bf s}} \: \omega \, f \left(\lambda_{\rm \bf s}-2 f+r f' \right)}\nonumber \\
&&  \hspace{0.9cm}+\frac{2 r}{\lambda_{\rm \bf s}\, f \left(\lambda_{\rm \bf s}-2 f+r f' \right)}{\biggl [}\left[r^2 \omega ^2+2 f^2-f \left(\lambda_{\rm \bf s}+r f'\right)\right]H_T+r f \left(2 f+r f'\right) H_T'+r^2f^2 H_T''{\biggr ]}\nonumber \\
&&  \hspace{0.9cm} +\frac{2 r^2 \left(f f_r'+f'f_r\right)}{\sqrt{\lambda_{\rm \bf s}}\left(\lambda_{\rm \bf s}-2 f+r f' \right)}. 
\end{eqnarray}
On the other hand the gauge invariant KI vector master field is 
\begin{equation}\label{KI:vectorPhi}
\Phi^{\rm (v)}_{\omega \ell_{\rm \bf v}}=\frac{i f}{\sqrt{\lambda_{\rm \bf v}} \:\omega }\left(\sqrt{\lambda_{\rm \bf v}} \:h_r+r h_T' \right).
\end{equation}

The KI master variables have the asymptotic expansion,
 \begin{equation}\label{KIasymp}
\Phi^{(j)}_{\omega_j \ell_j}{\bigl |}_{z\to 0}  \sim \Phi_0+\Phi_1 \,\frac{L}{r} +\cdots  \qquad  \hbox{where} \quad  j=\{ {\rm \bf s},{\rm \bf v}\}\,.
\end{equation}

 The linear differential map $h_{ab}^{(j)}=h_{ab}^{(j)}\left( \Phi^{(j)} \right)$ that reconstructs the metric perturbations (in a given gauge) can be read from \eqref{pert:scalar} and \eqref{KI:scalarPhi} (scalar case) and from \eqref{pert:vector} and \eqref{KI:vectorPhi} (vector case), if we follow  \cite{Kodama:2003jz}. The requirement that these metric perturbations are asymptotically global AdS in the sense described in Section \ref{sec:AsymGlobalAdS} imposes the conditions \cite{Michalogiorgakis:2006jc,Dias:2011ss}:
   \begin{eqnarray}
&& \hbox{Scalar BC:}  \qquad \Phi_1 =-\frac{6M}{\lambda_{\rm \bf s}-2}\, \Phi_0,\label{KI:BCS} \\
&& \hbox{Vector BC:}  \qquad   \Phi_0=0\,.   \label{KI:BCV}
\end{eqnarray}

We can of course consider different asymptotic BCs. For example, past studies on (quasi)normal modes have considered the BC $\Phi_0=0$ for scalar modes, instead of \eqref{KI:BCS}. However, these BCs are not asymptotically global AdS, i.e. they do not preserve the boundary metric, as first observed in \cite{Michalogiorgakis:2006jc}. We thus we do not consider them. Other BCs that might be called asymptotically globally AdS were studied in \cite{Compere:2008us}, but turn out to lead to ghosts (modes with negative kinetic energy) and thus make the energy unbounded below \cite{Andrade:2011dg}.

%%%%%%%%%%%%%%%%%%%%%%%%%%%%%%%%%%%%%%%%%%%%%%%%%%%%
\subsection{Teukolsky {\it vs} Kodama-Ishibashi (Regge-Wheeler$-$Zerilli) \label{sec:KIvsTeuk}}
%%%%%%%%%%%%%%%%%%%%%%%%%%  

Equations \eqref{KI:scalarH} and  \eqref{KI:vectorH} express the KI scalar $\scalar(x,\phi)$ and vector $\vector(x,\phi)$ harmonics as a function of the scalar spherical harmonic $Y_\ell^m(x,\phi)$ defined in \eqref{KI:scalarH}.
The spin-wheighted spherical harmonics $S_{\ell m}^{(s)}(x ,\phi )$ used in the Teukolsky harmonic decomposition are also related to the scalar spherical harmonic $Y_\ell^m(x,\phi)$ through the differential map \cite{Newman:1966ub,Goldberg:1966uu}
   \begin{eqnarray}
  && S_{\ell m}^{(-2)}(x ,\phi )=S_{\ell m}^{(2)}(-x ,\phi )\,;  \nonumber \\ 
&&S_{\ell m}^{(2)}(x ,\phi )=\sqrt{\frac{(\ell -2)!}{(\ell +2)!}}\sqrt{\frac{(\ell +1)!}{(\ell -1)!}}\left[\left(1-x^2\right)\partial _x\left(\frac{1}{\sqrt{1-x^2}}\,S_{\ell m}^{(1)}(x,\phi )\right)- \frac{i}{\sqrt{1-x^2}}\,\partial _{\phi }S_{\ell m}^{(1)}(x,\phi )\right]\!,  \nonumber \\ 
&& S_{\ell m}^{(1)}(x ,\phi )=\sqrt{\frac{(\ell -1)!}{(\ell +1)!}} \left[\sqrt{1-x^2}\, \partial _xY_{\ell m}(x,\phi )-\frac{i}{\sqrt{1-x^2}} \partial _{\phi }Y_{\ell m}(x,\phi )\right].\label{SpinWeightSphereH} 
  \end{eqnarray}
These (regular) harmonics obey the angular  Teukolsky equations \eqref{eqs:s-2}-\eqref{eqs:s+2}  with $a=0$ and  $\lambda=\ell(\ell+1)-2$.   

The two angular maps just described provide the starting point  to bridge the KI and Teukolsky formalisms. We now need the radial map.
The metric perturbations of the two formalisms must be the same (modulo gauge transformations), i.e. 
\begin{equation}\label{hKIvshTeuk}
h_{\mu\nu}^{\rm \bf s}+h_{\mu\nu}^{\rm \bf v}=h_{\mu\nu}^{\rm IRG}+h_{\mu\nu}^{\rm ORG} \,,
\end{equation}
where recall that the KI metric on the LHS is given by \eqref{pert:scalar},  \eqref{KI:scalarH}, \eqref{pert:vector} and  \eqref{KI:vectorH}. On the other hand, the Teukolsky metric on the RHS is given by \eqref{HertzIRG}, \eqref{HertzORG}, \eqref{HertzAnsatz} and \eqref{SpinWeightSphereH}. 

Fix the LHS  of \eqref{hKIvshTeuk} to have fixed values of KI quantum numbers $\ell_{\rm \bf s}$ and  $\ell_{\rm \bf v}$. Then, the most natural expectation is that such a KI perturbation is described by a (possibly infinite) sum, in the quantum number $\ell$, of Teukolsky harmonics (the background is spherically symmetric so we can set wlog $m=0$ in our discussion; see below).
In fact, a mode by mode analysis (using properties of internal products) reveals that $\ell_{\rm \bf s}=\ell_{\rm \bf v}=\ell$ (with no sum involved). This simplifies considerably the construction of the map.

Take \eqref{hKIvshTeuk} with the identification $\ell_{\rm \bf s}=\ell_{\rm \bf v}=\ell$  with integer $\ell\geq 2$. The later inequality requires a discussion before proceeding. 
The KI formalism describes all scalar modes with integer $\ell_{\rm \bf s}\geq 0$ (where $\ell_{\rm \bf s}=0$ are perturbations that just shift  the mass of the solution and $\ell_{\rm \bf s}=1$ is a pure gauge mode), and all vector modes with integer $\ell_{\rm \bf v}\geq 1$ (where $\ell_{\rm \bf v}=1$ are perturbations that generate just a shift in the angular momentum of the solution)  \cite{Kodama:2003jz,Dias:2013hn}.  However, the Teukosky quantum number is constrained to be an integer $\ell\geq |s|=2$, so the Teukolsky formalism is blind to the modes that generate deformations in the mass and angular momentum of the geometry  \cite{Wald,WaldL0L1}. The simplest way to confirm this is to note that the map \eqref{SpinWeightSphereH} would be trivial for the ``$\ell=0,1$" modes.

In these conditions, each metric component on the LHS of \eqref{hKIvshTeuk} is proportional to the spherical harmonic $Y_\ell^m(x)$, or to its first derivative, or to a linear combination of both such contributions. The same applies to the RHS of \eqref{hKIvshTeuk}. Matching all the coefficients of these angular contributions we can find the radial KI metric functions $\{ f_{ab},\,f_a,\,H_T,\,H_L\}$, $\{ h_a,\,h_T\}$ that describe the Teukolsky perturbations. Finally, inserting these radial KI functions into  \eqref{KI:scalarPhi} and \eqref{KI:vectorPhi}  we express the KI variables $\Phi^{({\rm s}, {\rm v})}_{\omega \ell}(r)$ as a function of the radial Teukolsky functions $R^{(s)}_{\omega \ell m}(r)$ and their first derivative. Note that the above KI metric functions are in a particular gauge induced by the Hertz construction. However, \eqref{KI:scalarPhi} and \eqref{KI:vectorPhi}, and thus the associated final map, are gauge invariant.
To write $\Phi^{({\rm s}, {\rm v})}_{\omega \ell}(r)$ as a function of a single Teukolsky function, e.g. $R^{(2)}_{\omega \ell m}(r)$, we further use  the radial Starobinsky-Teukolsky identity \eqref{StaroTeukRad} (which, recall, expresses $R^{(-2)}_{\omega \ell m}(r)$ as a function of $R^{(2)}_{\omega \ell m}(r)$ and its first derivative), and the equation of motion \eqref{eqs:s+2} for  $R^{(-2)}_{\omega \ell m}(r)$. We finally end up with the desired gauge invariant map between the KI scalar and vector  master fields $\Phi _{\omega  \ell  m}^{({\rm s}, {\rm v})}$ and the Teukolsky radial function $R^{(2)}_{\omega \ell m}$ (and its derivative):\begin{eqnarray}
  &&  \hspace{-1cm}\Phi _{\omega  \ell}^{(\rm \bf s)}=\frac{1}{ \mathcal{K}_{\rm st}\mathcal{C}_{\rm st} }\frac{\sqrt{(\ell +2)!}}{\sqrt{(\ell -2)!} }\:\frac{r^2}{2 f (\lambda  r+6M)}\left(\mathcal{K}_{\rm st} \mathcal{C}_{\rm st} \frac{(\ell -2)!}{(\ell +2)!}+\frac{(\ell +2)!}{(\ell -2)!}+12 \,i\, M \omega \right) \nonumber \\ 
  &&  \hspace{0.3cm} \times {\biggl \{ } r f \left[ \left(2 i \omega +f'\right) (\lambda  r+6M)-2 \lambda  f\right] \left(R_{\omega \ell}^{(2)}\right)' \nonumber \\ 
&&\hspace{0.9cm} +{\biggl[} 4 f^3-2 f^2 \left( \lambda +4-2 i r \omega +r f' \right)
     +r^2 \left(\lambda +2+r f'\right) \left(f'^2+3 i \omega  f'-2 \omega^2 \right) \nonumber \\
&&\hspace{1.3cm} +f \left[(\lambda +2-2 i \omega  r)^2+4r \omega (2 r \omega  +i)-r f'\left(\lambda -2+8 i r \omega +2 r f'\right)\right]  {\biggr ]}  R_{\omega \ell}^{(2)}    {\biggr \}},    \label{scalarPhiTeuk}   \\ 
  &&  
  \hspace{-1cm}\Phi _{\omega  \ell}^{(\rm \bf v)}= \frac{i}{\mathcal{K}_{\rm st} \mathcal{C}_{\rm st} }\frac{\sqrt{(\ell +2)!}}{ \sqrt{(\ell -2)!}}\: \frac{r}{4f}\left(-\mathcal{K}_{\rm st} \mathcal{C}_{\rm st} \frac{ (\ell -2)!}{ (\ell +2)!}-\frac{ (\ell +2)!}{ (\ell -2)!}+12 \,i\, M \omega \right)
   \nonumber \\ 
  &&  \hspace{0.4cm} \times {\biggl \{ }  \left[f \left(\lambda +2-4 i r \omega -2 r f'\right)+r^2 \left(f'^2+3 i \omega  f'-2 \omega ^2\right)\right]R_{\omega \ell}^{(2)} \nonumber \\
&&\hspace{0.9cm}  +r f \left(r f'-2 f+2 i \omega  r\right) \left(R_{\omega \ell}^{(2)}\right)'  {\biggr \}}.\label{vectorPhiTeuk} 
  \end{eqnarray}
In these expressions, $\lambda=\ell(\ell+1)-2$ is the spin-weighted eigenvalue for $a=0$ and $M$ is the mass parameter  of the black hole. Note that for $a=0$, the case we are discussing in this and following sections, the background spacetime is spherically symmetric. Consequently, the radial Teukolsky equations and solutions are independent of the azimuthal quantum number $m$.  Therefore, henceforth we droped the associated subscript in $R^{(2)}_{\omega \ell m}$ in the maps  \eqref{scalarPhiTeuk}-\eqref{vectorPhiTeuk} (and henceforward). The Starobinsky-Teukolsky  angular and radial constants ${\mathcal K}_{\rm st}$ and  ${\mathcal C}_{\rm st}$ are given by  \eqref{STconstC}-\eqref{STconstCf}, which for $a=0$ boil down to \footnote{We ask the reader to revisit the discussion leading to \eqref{STconstCf}. For that reason, and without any loss except compactness, we prefer to present our forthcoming results without explicitly giving the expression for $\mathcal{C}_{\rm st}$.}
\begin{equation}
\mathcal{K}_{\rm st}=\frac{ (\ell +2)!}{ (\ell -2)!}\,, \qquad 
 \mathcal{C}_{\rm st}=\mathcal{K}_{\rm st} - 12 \, i \, M \, \omega.
 \label{a0:STconst}
\end{equation}
To check our matching we explicitly verify that our KI master fields obey the KI master equations \eqref{KImaster} when $R^{(2)}_{\omega \ell }(r)$ satisfies the radial Teukolsky equation \eqref{eqs:s+2}, with $\lambda=\ell(\ell+1)-2$. Note that the map \eqref{scalarPhiTeuk}-\eqref{vectorPhiTeuk}  is valid also in the asymptotically flat limit $L\to \infty$.

To have the complete map between the Teukolsky and KI formalism we still need to discuss the relation between the asymptotically global AdS KI BCs \eqref{KIasymp}-\eqref{KI:BCV} and the global AdS Teukolsky BCs \eqref{TheBCs}-\eqref{TheBCsV}. The later, for $a=0$, simply reduce to  \begin{eqnarray}
&&  1) \quad A_-^{(2)}=-i\,\eta \, A_+^{(2)}, \qquad \eta=\eta_{\rm \bf s}= -L \omega  \left(1+\frac{\lambda }{\lambda -2\left(L^2 \omega ^2-1\right)}\right), \label{TheBCsSa0}  \\ 
&&  2) \quad A_-^{(2)}=-i\,\eta \, A_+^{(2)}, \qquad   \eta=\eta_{\rm \bf v}=\frac{\lambda }{2 L \omega }-L \omega \,. \label{TheBCsVa0}
\end{eqnarray}
It follows from \eqref{FrobR} and \eqref{TheBCs} that asymptotically  
\begin{equation}\label{FrobR2}
 R_{\omega \ell}^{(2)}{\bigl |}_{r\to \infty}\sim  A_{+}^{(2)}\,\frac{L}{r}\left[ 1-i\, \eta \,\frac{L}{r}+\frac{1}{2} \left(\ell ^2+\ell -4-L^2 \omega ^2\right) \frac{L^2}{r^2} \right]+\mathcal{O}\left(  \frac{L^4}{r^4}  \right)\,,
\end{equation}

Consider first the scalar case  described by  \eqref{scalarPhiTeuk}. Choose  the BC to be such that $\eta=\eta_{\rm \bf s} $ as defined in \eqref{TheBCsSa0}. In these conditions, inserting \eqref{FrobR2} into the scalar map \eqref{scalarPhiTeuk}  and taking its asymptotic expansion we find that it reduces exactly to the KI expression \eqref{KIasymp}, $\Phi^{(\rm \bf s)} \sim \Phi_0+\Phi_1 \,\frac{L}{r}$, with 
\begin{eqnarray}\label{scalarPhiTeukBC}
&&\Phi_1 =-\frac{6M}{\ell (\ell +1)-2}  \Phi_0, \\
&&\Phi _0=\frac{L \, A_+^{(2)}}{2 C_{\rm st}}\:\frac{ \sqrt{(\ell -2)!}}{\sqrt{(\ell +2)!}}\:\frac{ \frac{ (\ell +2)!}{ (\ell -2)!}-12 i M \omega }{\ell (\ell +1)- 2L^2 \omega ^2}\left(C_{\rm st} +\frac{ (\ell +2)!}{ (\ell -2)!}+12 i M \omega \right), \nonumber 
%&&\Phi _0=\frac{\sqrt{(\ell +2)!}}{\sqrt{(\ell -2)!}} \frac{L\,A_+^{(2)}}{ \ell (\ell +1)- 2L^2 \omega ^2} \nonumber
\end{eqnarray}
which  matches  the global AdS KI BC \eqref{KI:BCS} for scalar modes once we use $\lambda_{\rm \bf s}=\ell (\ell +1)$. We see this as one of the most non-trivial tests of our calculations.

Next,  take the vector case described by  \eqref{vectorPhiTeuk}. This time select  the BC $\eta=\eta_{\rm \bf v} $  defined in \eqref{TheBCsVa0}. Plug \eqref{FrobR2} into the vector map \eqref{vectorPhiTeuk}  and take its asymptotic expansion.  We get the KI expression \eqref{KIasymp} with 
\begin{equation}\label{vectorPhiTeukBC}
\Phi_0 =0, \qquad
\Phi _1=\frac{L \, A_+^{(2)}}{8\omega  C_{\rm st}}\: \frac{\sqrt{(\ell -2)!}}{\sqrt{(\ell +2)!}} \left(\frac{(\ell +2)!}{(\ell -2)!}+12 i M \omega \right) \left(C_{\rm st}+\frac{(\ell +2)!}{(\ell -2)!}-12 i M \omega \right),
% \Phi _1= -\frac{L A_+^{(2)}}{4 \omega }\frac{\sqrt{(\ell -2)!}}{\sqrt{(\ell +2)!}} \left(\frac{(\ell +2)!}{(\ell -2)!}+12 i M \omega \right)
\end{equation}
 which is the global AdS KI BC \eqref{KI:BCV} for vector modes. 

So one of the Teukolsky Robin BCs selects the scalar sector of KI perturbations and the other selects the vector KI sector. This is the simplest map we could have predicted!

To sum this section, we found the differential map between the Teukolsky/KI variables and BCs. For scalar modes, the differential map is given by \eqref{scalarPhiTeuk}  with the global AdS Teukolsky BC \eqref{TheBCsSa0} mapping into the scalar KI BC \eqref{KI:BCS} via \eqref{scalarPhiTeukBC}. For vector modes, the differential map is instead 
 \eqref{vectorPhiTeuk}, and the global AdS Teukolsky BC \eqref{TheBCsVa0} maps into the vector KI BC  \eqref{KI:BCV} through \eqref{vectorPhiTeukBC}.
By continuity, when we turn on the rotation $a$, we can say that the Teukolsky BC \eqref{TheBCs},\eqref{TheBCsS} generates the ``rotating scalar modes", while  the BC \eqref{TheBCs},\eqref{TheBCsV} selects the ``rotating vector modes".

%%%%%%%%%%%%%%%%%%%%%%%%%%%%%%%%%%%%%%%%%%%%%%%%%%%%
\section{Global AdS (quasi)normal  modes  \label{sec:QNmodes}}
%%%%%%%%%%%%%%%%%%%%%%%%%%

The normal modes of global AdS can be studied using the Teukolsky equations and the boundary conditions  \eqref{TheBCs}-\eqref{TheBCsV}.  We take the opportunity to find these normal mode frequencies since the scalar modes are not explicitly derived in the literature. We also revisit, from a Teukolsky perspective, the quasinormal mode spectrum of the global AdS-Schwarzschild black hole. 

Introducing the differential operator definitions \eqref{diffOp},  the $s=+2$ the Teukolsky equations \eqref{eqsF:s+2}  read
\begin{eqnarray}\label{eqsF:s+2}
&&   \partial _{\chi }\left(\Delta _{\chi }\partial _{\chi } S_{\omega \ell m}^{(2)} \right)+ \left[-\frac{\left(K_{\chi }-\Delta _{\chi }'\right)^2}{ \Delta _{\chi }}+ \left(\frac{6\chi ^2}{L^2}-4K_{\chi }'+\Delta _{\chi }''\right)+\lambda \right]S_{\omega \ell m}^{(2)}=0   \,, \nonumber \\
 && \partial _r\left(\Delta _r\partial _r R_{\omega \ell m}^{(2)}\right)+\left[\frac{\left(K_r+i \Delta _r'\right)^2}{ \Delta _r}+\left(\frac{6 r^2}{L^2}-4i K_r'+ \Delta _r''\right)-\lambda \right]R_{\omega \ell m}^{(2)}=0\,.
 \end{eqnarray}
It follows from \eqref{CMtoBL} that in the static case it is appropriate to work in the coordinate system $\{t,r,x,\phi\}$ where $\chi=a \, x$. 
In these conditions equations \eqref{eqs:s+2} with $a=0$ reduces to 
\begin{eqnarray}\label{eqsa0:s+2}
&&   \partial _x\left[\left(1-x^2\right)\partial _x \,S_{\omega \ell m}^{(2)}(x)\right]+\left[\lambda -2-\frac{(m+2 x)^2}{1-x^2}\right] S_{\omega \ell m}^{(2)}(x)=0\,, \nonumber \\
 && \partial _r\left(\Delta _r\partial _r\,R_{\omega \ell}^{(2)}(r)\right)+ \left[\frac{\left(\omega  r^2-i \Delta _r'\right)^2}{ \Delta _r}+2\left(1+\frac{9 r^2}{L^2}\right)+8 i r \omega -\lambda \right]R_{\omega \ell}^{(2)}(r)=0\,. \end{eqnarray}
The spin weighted spherical harmonic is independent of the mass parameter $M$ and cosmological radius $L$, and can be found analytically. Next, we discuss in detail the regularity analysis  that leads to the solution \eqref{SpinWeightSphereH}. 

The regular solution at the north pole $x=1$ is  ($_2F_1$ is the standard Hypergeometric function)
 \begin{eqnarray}\label{AngSola00}
&&S_{\omega \ell m}^{(2)}=(1-x)^{\frac{|m+2|}{2}} (1+x)^{\frac{|m-2|}{2}} \,  _2F_1\left(\frac{1}{2} \left(\tilde{a}-\tilde{b}\right),\frac{1}{2} \left(\tilde{a}+\tilde{b}\right),|m+2|+1,\frac{1-x}{2}\right)\,, \nonumber \\
&&
 \hbox{with}\quad \tilde{a}=|m-2|+|m+2|+1\,, \quad \tilde{b}=\sqrt{4 \lambda +9}\,.
\end{eqnarray}
This solution diverges at the south pole  $x=-1$ as a positive power of $(1+x)^{-1}$ (or as $\ln(1+x)$ in the special case of $m=2$) unless we  quantize the angular eigenvalue and quantum numbers as
 \begin{equation}\label{condAngN}
 \lambda=\ell (\ell +1)-2\,, \qquad \hbox{with}\qquad \ell=2,3,4,\cdots\,,\quad  |m|\leq \ell\,,
\end{equation}
where we have introduced the quantum number $\ell$ with $\ell-{\rm max}\{|m|,|s|=2\}$ giving the number of zeros of the eigenfunction along the polar direction. 
The regular spin $s=2$ spherical harmonic that solves the angular equation \eqref{eqsa0:s+2}
is finally
  \begin{eqnarray}\label{AngSola0}
&&\hspace{-0.5cm} S_{\omega \ell m}^{(2)}=(1-x)^{\frac{|m+2|}{2}} (1+x)^{\frac{|m-2|}{2}} \,   \\
&&
\hspace{0.7cm}\times \,  _2F_1\left(\frac{1}{2} (-2 \ell +|m-2|+|m+2|),\frac{1}{2} (2+2 \ell +|m-2|+|m+2|),|m+2|+1,\frac{1-x}{2}\right).\nonumber
\end{eqnarray}
with the quantum numbers $\ell,m$ constrained by the conditions \eqref{condAngN}.
This harmonic is valid both for the global AdS-Schwarzchild and global AdS backgrounds since the angular equation is independent of $M$. Using the relation between the Hypergeometric function and the Associated Legendre polynomial we can rewrite \eqref{AngSola0} as \eqref{SpinWeightSphereH}.

To study the (quasi)normal modes of these backgrounds we now need to study the radial equation \eqref{eqsa0:s+2}. 
Since this equation depends on the mass parameter $M$ we need to study the cases $M=0$ and $M>0$ separately. In the next subsection  we first find the normal modes of global AdS and we study the quasinormal modes  of global AdS-Schwarzchild in subsection \eqref{sec:QN}. 

%%%%%%%%%%%%%%%%%%%%%%%%%%%%%%%%%%%%%%%%%%%%%%%%%%%%
\subsection{Normal  modes of global AdS \label{sec:QNGlobalAdS}\label{sec:normalM}}
%%%%%%%%%%%%%%%%%%%%%%%%%%

In the global AdS background ($M=0$), the radial Teukolsky equation \eqref{eqsa0:s+2} has an exact solution. The solution that is regular at the origin ($r=0$) is 
 \begin{equation}\label{AngRada0}
R_{\omega \ell}^{(2)}= A_0 \left(1-\frac{i\, r}{L}\right)^{\frac{1}{2} (L \omega -2)}\!\!\! \left(1+\frac{i\, r}{L}\right)^{-\frac{1}{2} (L \omega +2 \ell )}\!\! \!\left(\frac{r}{L}\right)^{\ell } \, _2F_1\left(\ell -1,\ell +1+L \omega,2 (\ell +1),\frac{2 r}{r-i\, L}\right).
\end{equation}
where $A_0$ is an arbitrary amplitude.
Asymptotically this solution behaves as
\begin{eqnarray}\label{AngRada0BC}
&& \hspace{-1.2cm}R_{\omega \ell}^{(2)}{\biggl |}_{r\to \infty}\!\! \sim 
-A_0 e^{-\frac{1}{2} i \pi  (L \omega -3 \ell )}{\biggl [}-i\frac{ L}{r} \, _2F_1(\ell -1,\ell +1+L \omega ,2 (\ell +1),2) \nonumber \\
&& \hspace{2cm}+\frac{L^2 }{r^2}{\biggl (}(L \omega +\ell -1) \, _2F_1(\ell -1,\ell +1+L \omega ,2 (\ell +1),2) \nonumber \\
&&   \hspace{3cm}+\frac{(\ell -1) (L \omega +\ell +1) }{\ell +1}\, _2F_1(\ell ,\ell +2+L \omega ,2 \ell +3,2){\biggr )}{\biggr ]}\!+\mathcal{O}\left(\frac{L^3}{r^3}\right)\!.
\end{eqnarray}
Comparing this decay with \eqref{FrobR} we can read the expressions for two amplitudes $A_{+}^{(2)}$ and $A_{-}^{(2)}$. These amplitudes are {\it \`a priori} independent but the requirement that the perturbation is asymptotically global AdS constrains them to be related by the BCs \eqref{TheBCsVa0}.
These BCs quantize the frequencies of the perturbations that can fit in the global AdS box, respectively, as
 \begin{eqnarray}
&& 1) \quad \hbox{Scalar normal modes of global AdS:} \quad \omega L=1+\ell+2p  \,, \label{SnormalM}
\\ 
&& 2) \quad \hbox{Vector normal modes of global AdS:} \quad \omega L=2+\ell+2p \,,\label{VnormalM}   \end{eqnarray}
where the non-negative integer $p$ is the radial overtone that gives the number of nodes along the radial direction and recall that $\ell\geq 2$ is an integer.  The frequencies \eqref{SnormalM} and \eqref{VnormalM} describe, respectively, the scalar  and vector normal mode frequencies of global AdS. Note that we are adopting the standard KI classification of the scalar/vector perturbations, in the sequence of our conclusions of the previous section. Without any surprise, the frequencies \eqref{SnormalM} and \eqref{VnormalM} precisely agree with the normal mode frequencies of global AdS that we obtain when we solve the KI master equation \eqref{KImaster} subject to the BCs \eqref{KI:BCS} , i.e.  $\Phi_1 =0$ in the scalar case \cite{Michalogiorgakis:2006jc,Dias:2011ss} and $\Phi_0=0$ in the vector case \cite{Natario:2004jd,Dias:2012tq}.

%%%%%%%%%%%%%%%%%%%%%%%%%%%%%%%%%%%%%%%%%%%%%%%%%%%%
\subsection{Quasinormal  modes of global AdS-Schwarzchild \label{sec:QNSchwAdS}\label{sec:QN}}
%%%%%%%%%%%%%%%%%%%%%%%%%%

In this subsection we study some properties of the gravitational quasinormal mode (QNM) spectrum of the global AdS-Schwarzschild black hole (GAdSBH). In the AdS$_4$/CFT$_3$ duality, this spectrum is dual to the thermalization timescales of the perturbed thermal states of the CFT$_3$ living on the sphere, as discussed in \cite{Michalogiorgakis:2006jc} (following the detailed analysis of the AdS$_5$/CFT$_4$ case presented in  \cite{Friess:2006kw}).  We focus our attention in the low-lying QNMÕs (small radial overtone $p$ and harmonic $\ell$) because they are expected to dominate the late-time behavior of the time evolution.

Many properties of this gravitational QNM spectrum were already studied with some detail in the past. The low-lying KI vector QNMs with global AdS BCs were discussed in \cite{Cardoso:2001bb,Berti:2003ud,Michalogiorgakis:2006jc}. The asymptotic behavior of these vector modes for large overtone were further analyzed in \cite{Cardoso:2001bb}-\cite{Siopsis:2007wn} (see footnote \ref{foot}). On the other hand, the low-lying KI scalar QNMs with global AdS BCs were studied in \cite{Michalogiorgakis:2006jc} (see also \cite{Friess:2006kw}). Finally, the asymptotic behavior of the vector/scalar QNMs for large harmonic $\ell$ was found in a WKB analysis in \cite{Dias:2012tq}.

Our results agree with the vector results of Cardoso-Lemos \cite{Cardoso:2001bb} and with the results of  Michalogiorgakis-Pufu \cite{Michalogiorgakis:2006jc}. Our conclusions and presentation contribute to complement these previous analysis mainly by plotting the QNM spectrum as a function of the horizon radius, and not just a few points of the spectrum. Our discussion will always focus on the parameter space region of $r_+/L$ where the relevant physics is and/or where the spectrum varies the most. Given the $t-\phi$ symmetry of the GAdSBH, the QNM frequencies always come in trivial pairs of $\{\omega,-\omega^*\}$. We just plot the element of the pair with positive real frequency.

To find the QNM spectrum we solve the Teukolsky radial equation \eqref{eqsa0:s+2} numerically subject to the asymptotically global AdS BCs, namely, \eqref{TheBCsSa0} in the scalar case and  \eqref{TheBCsVa0} in the vector case. We use spectral methods to solve the numerical problem, which uses a Chebyshev discretization of the grid. We work with the compact radial coordinate $0\leq y\leq 1$ and with the new radial function $q_j$ defined as
\begin{eqnarray}
&& \hbox{Vector:} \:\:\: y=\sqrt{1-\frac{r_+}{r}},  \qquad R_{\omega \ell}^{(-2)}= \left(1-y^2\right)y^{2-\frac{i \,\omega L}{2\pi  T_H}} q_{\rm v}\,; \\
&& \hbox{Scalar:} \:\:\: y=1-\frac{r_+}{r},  \qquad R_{\omega \ell}^{(-2)}= (1-y) y^{1-\frac{i \omega  L}{4\pi  T_H}} \left(1-\frac{L^2}{r_+^2}\frac{ \left(1-16 \pi  T_H\left.r_+\right/L \right)}{6 \pi  T_H}(1-y)\right)^{-\frac{i \omega  L}{2}} q_{\rm s}.   \nonumber
 \end{eqnarray}
The horizon BC \eqref{bcH} translates  to a simple Neumann BC and the asymptotic BC  \eqref{TheBCsVa0}  yields a Robin BC, in the vector case. For the scalar case, both the horizon BC \eqref{bcH} and the asymptotic BC  \eqref{TheBCsSa0}  translate to a Robin BC relating $q_{\rm s}$ and its derivative.
In both cases, we get a generalized quadratic eingenvalue problem in the (complex) QNM frequencies $\omega$. We give the harmonic $\ell$ and run the code for several dimensionless horizon radius $r_+/L$.

To discuss the results consider first the vector QNM spectrum. We will either plot the imaginary part of the dimensionless frequency ${\rm Im}(\omega L)$ as a function of the real part ${\rm Re}(\omega L)$ or the QNM real/imaginary parts as a function of  the horizon radius in AdS units ($r_+/L$).

 \begin{figure}[ht]
\centering
\includegraphics[width=.47\textwidth]{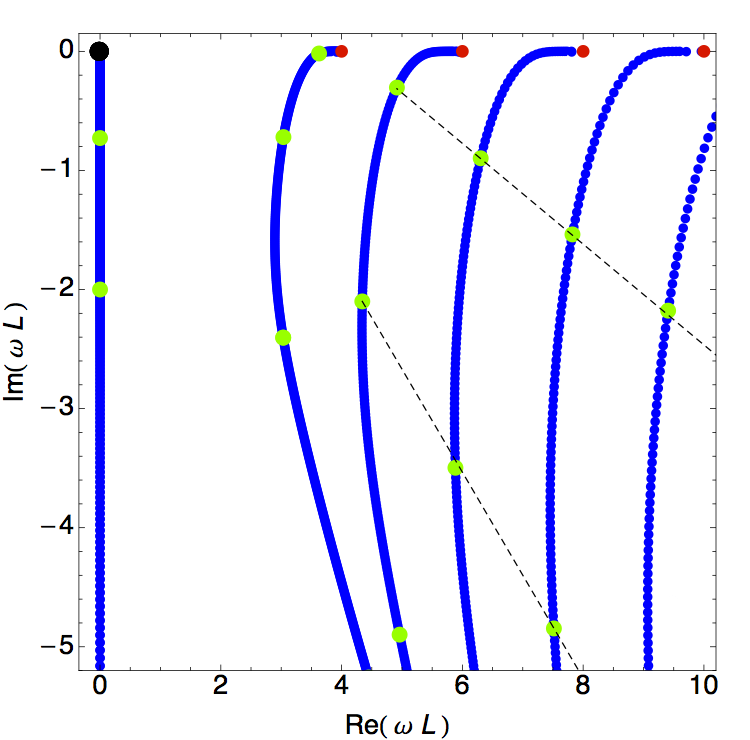}
\hspace{0.5cm}
\includegraphics[width=.47\textwidth]{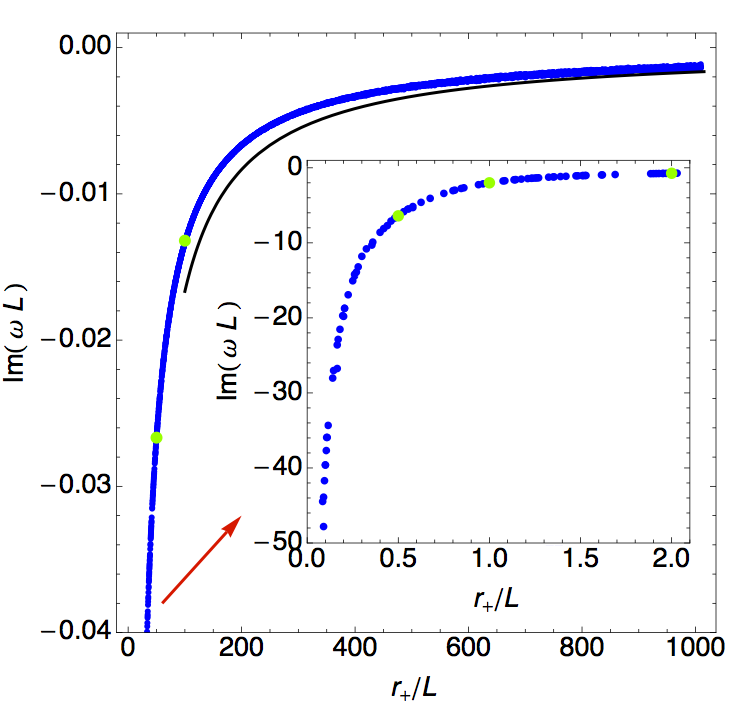}
\caption{{\it Left} panel: Hydrodynamic QNM (which ends in the black point) and the first four microscopic QNM curves (that start at the vector normal modes of AdS pinpointed as red dots) of the $\ell=2$ harmonic of the vector QNM spectrum of GAdSBH.
{\it Right} panel: Imaginary part of the hydrodynamic vector QNM  as a function of the horizon radius in AdS units. For large $r_+/L$, the data approaches the black curve which is the analytical prediction \eqref{hydroV}. The green dots in the hydrodynamic and in the lowest-lying ($p=0$) microscopic QNM curves are exactly the values taken from Table 2 of  \cite{Cardoso:2001bb}.
See text for detailed discussion of these plots.}\label{Fig:vectorQNovertones}
\end{figure}   

On the  {\it Left} panel of Fig. \ref{Fig:vectorQNovertones} we plot the first five low-lying QNMs of the $\ell=2$ vector harmonic. The points in the vertical line have pure imaginary frequencies and, as $r_+/L$ grows large, it approaches the black point $(0,0)$. They are the vector {\it hydrodynamic modes} since, in the limit $r_+\gg L $, they can be found solving the perturbed Navier-Stokes equation that describes the hydrodynamic regime of the CFT$_3$  on the sphere (the associated plasma is conformal, hence it has zero bulk viscosity and shear to entropy density ratio $\eta/s=1/(4\pi)$) \cite{Michalogiorgakis:2006jc,Friess:2006kw}.  
To leading order in the inverse of the horizon radius, this hydrodynamic computation yields the frequency  \cite{Michalogiorgakis:2006jc}
 \begin{equation}\label{hydroV}
\omega L {\bigl |}_{hydro}= i\,\frac{\ell^2+2\ell-1}{3}\,\frac{L}{r_+} + \mathcal{O}\left( \frac{L^2}{r_+^2}\right).   
 \end{equation}
Recall that the hydrodynamic regime requires that the perturbation wavelength is much larger than the thermal scale (the inverse of the temperature) of the theory. So this regime is achieved  when $r_+/L$ (and thus the temperature  $T_H L$) grows without bound and the perturbation frequency becomes arbitrary small. This is indeed what happens as we approach the black point  moving from bottom to top along the vertical line of Fig. \ref{Fig:vectorQNovertones}. On the {\it Right} panel of this figure we plot the (imaginary) frequency of the hydrodynamic mode as a function of    $r_+/L$ for $0<r_+/L<1000$. The black curve describes the analytical hydrodynamic curve \eqref{hydroV}. As predicted in \cite{Michalogiorgakis:2006jc}, we confirm that as $r_+/L$ grows the  black curve indeed approaches the numerical data.

\begin{figure}[t]
\centering
\includegraphics[width=.47\textwidth]{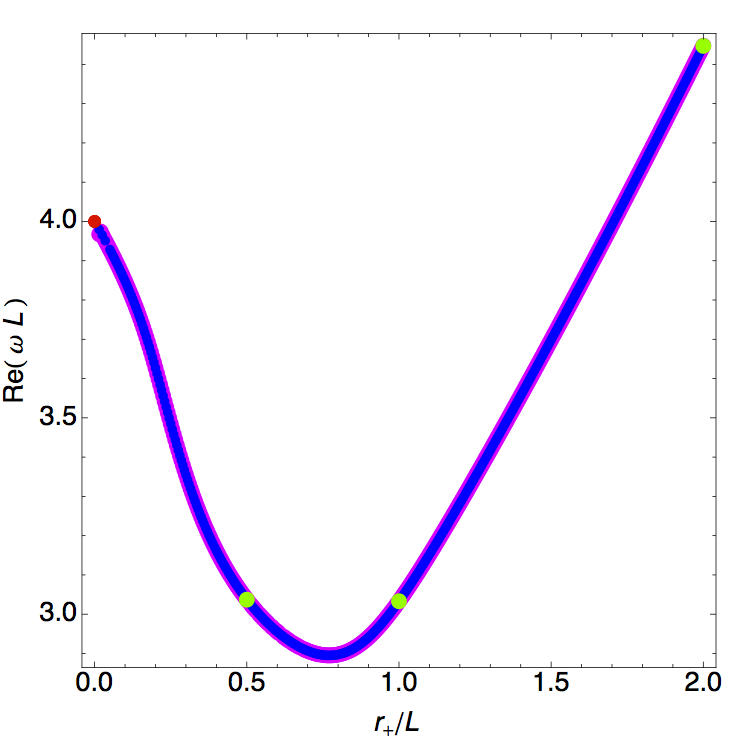}
\hspace{0.5cm}
\includegraphics[width=.47\textwidth]{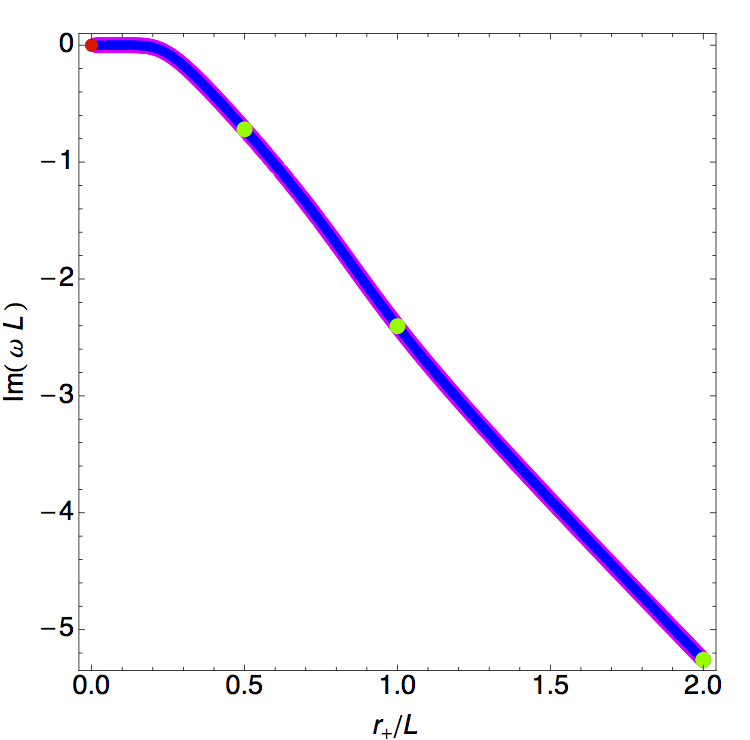}
\caption{Lowest-lying ($p=0$) microscopic vector QNM of the $\ell=2$ harmonic of GAdSBH. The green dots are exactly the values taken from Table 2 of  \cite{Cardoso:2001bb}. The blue (magenta) dots are obtained solving numerically the Teukolsky (KI) equations. (See text for detailed discussion of these plots).}\label{Fig:QNvector}
\end{figure}   

Returning to the {\it Left} panel of Fig. \ref{Fig:vectorQNovertones}, the red points describe  the first 4 radial overtones ($p=0,1,2,3$) of the $\ell=2$ vector normal modes of AdS; see \eqref{VnormalM}. The associated four vector QNMs in the {\it Left} panel of Fig. \ref{Fig:vectorQNovertones} are {\it microscopic modes} (as oppose to hydrodynamic) because as these curves move away from the red points, i.e. as $r_+/L$ (and $T_H L$) grows so does Im$(\omega L)$. So we never reach the hydrodynamic regime $\omega \ll T_H$ and we need the microscopic theory to describe them. This tower of overtones which are continuously connected to the normal modes of AdS are often said to form the main series or main sequence of the vector QNM spectrum.  In the plot we also pinpoint with green dots the points with  fixed $r_+/L=0.5$ and $r_+/L=1.0$ in each overtone curve (for the $p=0$ and microscopic mode curves we also identify the green mode with $r_+/L=2$). 
The green dots in the hydrodynamic and in the lowest-lying ($p=0$) microscopic QNM curves are also exactly the values taken from Table 2 of  \cite{Cardoso:2001bb}.
Modes with fixed  $r_+/L$ in the main sequence and for $p\geq 1$ scale linearly with $r_+/L$. To illustrate this property we connect with auxiliary dashed black straight line the two set of modes with $r_+/L=0.5,1$  that we have singled out. The points of the lowest-lying ($p=0$) QNM curve of the main sequence do not however fit in these lines as is visible in the plot, but apart from this curiosity, the $p=0$ overtone curve is similar to the higher overtone curves.

\begin{figure}[th]
\centering
\includegraphics[width=.47\textwidth]{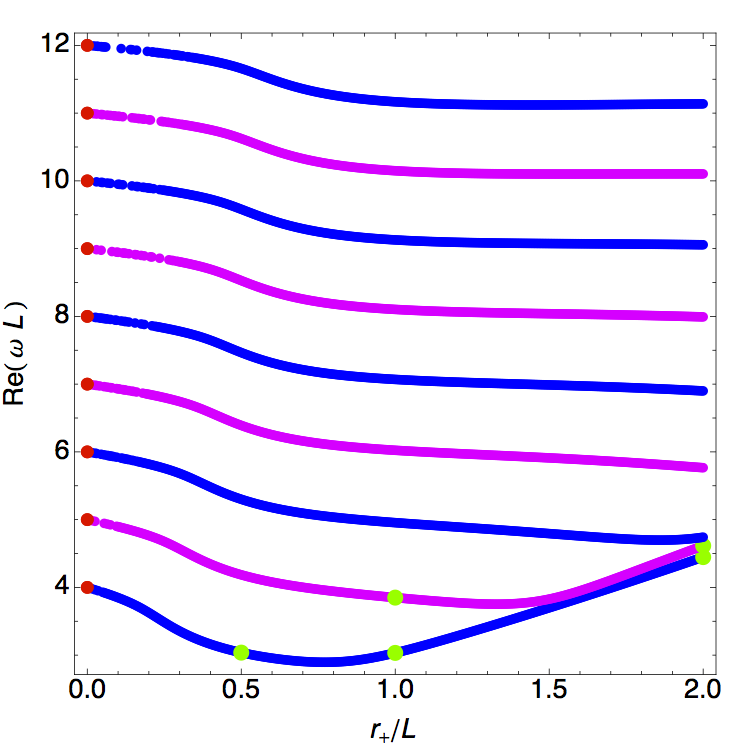}
\hspace{0.5cm}
\includegraphics[width=.47\textwidth]{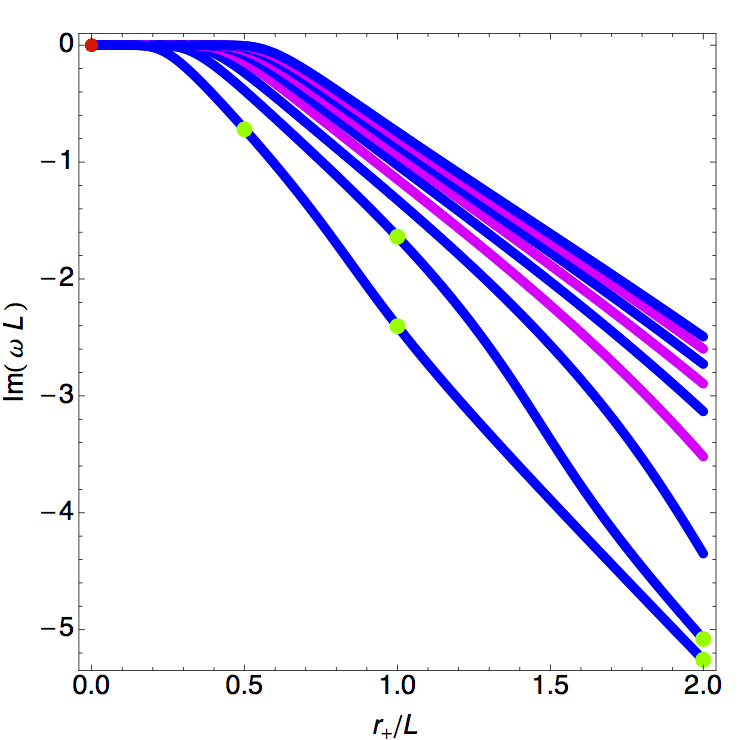}
\caption{Lowest-lying ($p=0$) microscopic vector QNMs of the first 8 harmonics of the GAdSBH. 
From bottom to top we have: $\ell=2,3,\cdots,10$. The red dots give the vector normal mode frequencies \eqref{VnormalM}. The green dots are exactly taken from Table 2 of  \cite{Cardoso:2001bb}.}\label{Fig:QNvector2}
\end{figure}   

Further properties of this lowest-lying ($p=0$) microscopic QNM curve are displayed in  Fig. \ref{Fig:QNvector}. 
The  {\it Left} ({\it Right}) panel plots the real (imaginary) part of the QNM  frequency as a function of the horizon radius for $r_+/L\leq 2$. The curve starts at the vector AdS normal mode frequency (red point) and the frequency stays very close to the real axis for $r_+/L< 0.2$. The real (imaginary) part keeps (increasing) decreasing monotonically  for larger values of $r_+/L$ (e.g. for $r_+=100 L$ one has $\omega L\sim 266.384- 184.959\, i$ \cite{Cardoso:2001bb}).  The same 3 green points for  $r_+/L=0.5,1,2$ of Fig. \ref{Fig:vectorQNovertones} are also plotted here. As an extra check of our numerics, in addition to the blue points, which are obtained solving numerically the Teukolsky system with BC \eqref{TheBCsVa0}, we also display with magenta points the numerical solution of the KI system with BC \eqref{KI:BCV}. In the sequence of the map constructed in Section \ref{sec:mapKIvsTeuk} these two curves have to overlap. This is indeed the case (the magenta points are drawn larger to be visible in the plot).

To analyze the evolution of the vector QNM spectrum as the vector harmonic changes, in Fig. \ref{Fig:QNvector2} we plot the same information as in Fig. \ref{Fig:QNvector}, and in addition the lowest-lying ($p=0$) microscopic QNM of the next seven vector harmonics $\ell$. More concretely, from bottom to top we have the harmonics $\ell=2,3,\cdots,10$. From the {\it Left} panel, we conclude that there are regions in the parameter space (i.e. windows in the range of $r_+/L$) where the real part of the frequency spectrum is in a first approximation isospectral (i.e. difference between consecutive harmonics at constant $r_+/L$ is approximately constant) but there are also others where the lowest harmonics $\ell=2,3$ spoil this property. From the {\it Right} panel, we see that the window of $r_+/L$ around the global AdS case ($r_+/L=0$; red dot (0,0)) where the imaginary part of the spectrum is approximately flat increases as the harmonic $\ell$ grows.
The green dots in the $\ell=2,3$ curves are exactly the values taken from Table 2 of  \cite{Cardoso:2001bb}.
 
   \begin{figure}[ht]
\centering
\includegraphics[width=.47\textwidth]{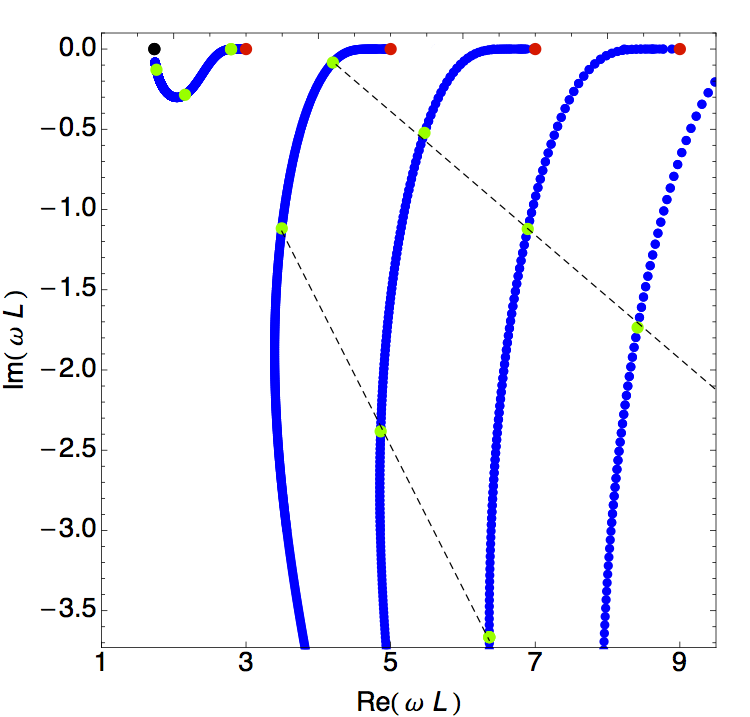}
\caption{   Hydrodynamic QNM (which ends at the black point) and the first three microscopic QNM curves (that start at the vector normal modes of AdS pinpointed as red dots) of the $\ell=2$ harmonic of the scalar QNM spectrum of GAdSBH. The green dots have exactly the values taken from Table 1 and 2 of \cite{Michalogiorgakis:2006jc} (we have added the $r_+=0.5$ green points in the $p\geq 1$ curves). (See text for detailed discussion of this plot).}\label{Fig:scalarQNovertones}
\end{figure}   

\begin{figure}[ht]
\centering
\includegraphics[width=.45\textwidth]{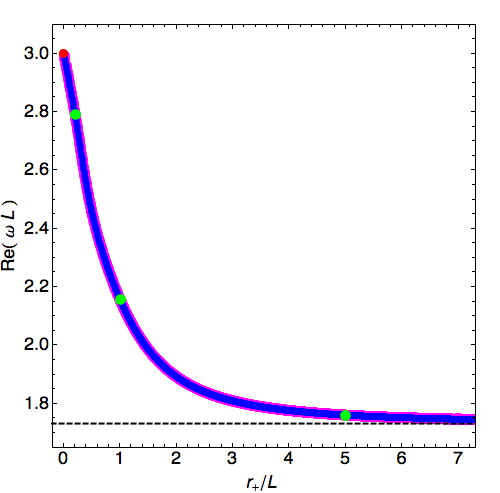}
\hspace{0.5cm}
\includegraphics[width=.47\textwidth]{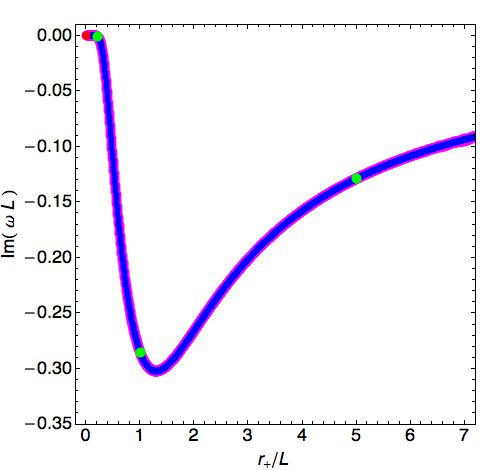}
\caption{Hydrodynamic scalar QNM of the $\ell = 2$ harmonic of GAdSBH. The green dots are exactly the values taken from taken from Table 1 and 2 of \cite{Michalogiorgakis:2006jc}.  The blue (magenta) dots are obtained solving numerically the Teukolsky (KI) equations. (See text for detailed discussion of these plots).}\label{Fig:QNscalar}
\end{figure}   

Consider now the scalar QNM spectrum of the GAdSBH. 
 In Fig. \ref{Fig:scalarQNovertones} we plot the first four low-lying QNMs of the $\ell=2$ scalar harmonic. In this case, the spectrum has no pure imaginary frequencies, and all the QNM curves are continuously connected to the scalar normal modes frequencies \eqref{SnormalM} of AdS here coloured as red dots for the overtones $p=0,1,2,3$. Clearly, the $p=0$ overtone curve on the left is a special curve. Indeed, in this case as $r_+/L$ (and $T_H L$) grows the imaginary part of the frequency first decreases but then it has a minimum after which it approaches zero (see the black point). That is, we approach the hydrodynamic regime $\omega \ll T_H$. This is thus the scalar hydrodynamic QNM. For the $p\geq 1$ the imaginary part of the frequency decreases monotonically as $r_+/L$: these are microscopic scalar QNMs. 
 Solving the linearized hydrodynamic equations on $R_t\times S^2$ for a conformal plasma,  
 \cite{Michalogiorgakis:2006jc} finds that to leading order the scalar hydrodynamic QNM is described by
 \begin{equation}\label{hydroS}
\omega L {\bigl |}_{hydro} = \frac{\sqrt{\ell(\ell+1)}}{\sqrt{2}}-i\,\frac{\ell^2+2\ell-2}{6}\,\frac{L}{r_+} + \mathcal{O}\left( \frac{L^2}{r_+^2}\right),   
 \end{equation} 
 and this fixes the black point in Fig. \ref{Fig:scalarQNovertones} when $r_+/L\to \infty$. This analytical result was already compared against numerical data for large radius in \cite{Michalogiorgakis:2006jc}. 
Much like in the vector case, in the main sequence ($p\geq 1$), microscopic modes with fixed  $r_+/L$ scale linearly with $r_+/L$. To illustrate this property we connect with auxiliary dashed black straight line the two set of modes with $r_+/L=0.2,\,0.5$. The green dots have exactly the values taken from Table 1 and 2 of \cite{Michalogiorgakis:2006jc} (we have added the $r_+=0.5$ green points in the $p\geq 1$ curves).

\begin{figure}[ht]
\centering
\includegraphics[width=.47\textwidth]{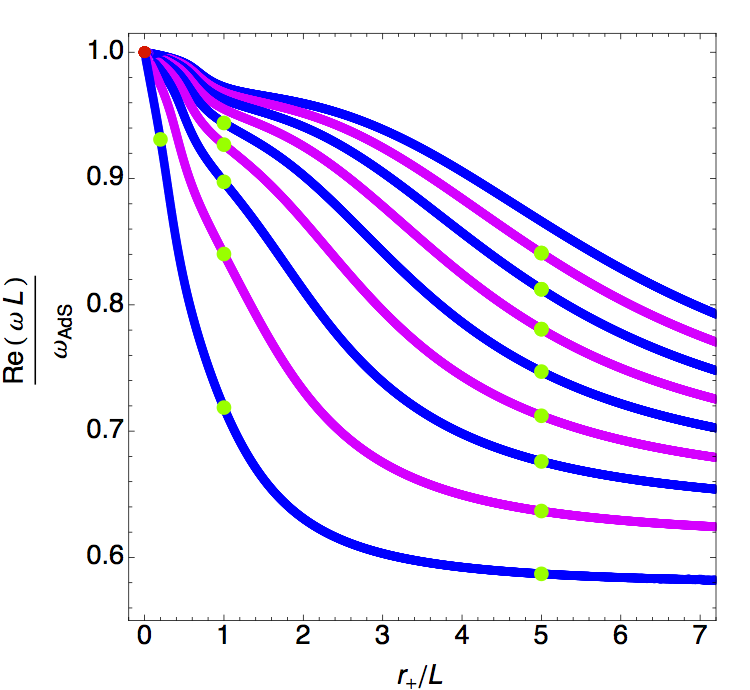}
\hspace{0.5cm}
\includegraphics[width=.47\textwidth]{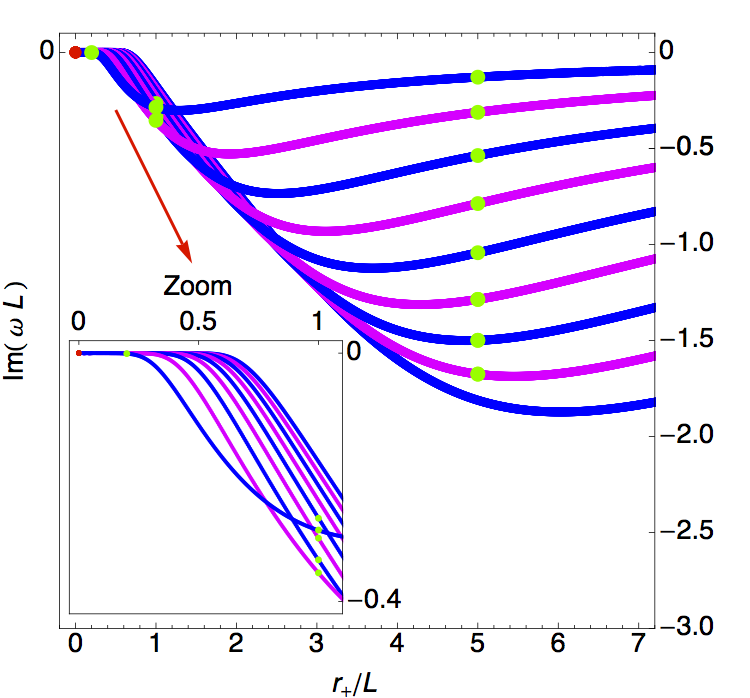}
\caption{
Hydrodynamic scalar QNMs of the first 8 harmonics of the GAdSBH. Viewing from the right side of the plots, from bottom to top we have: $\ell = 2,3,ááá ,10$. The red dots give the scalar normal mode frequencies  \eqref{SnormalM} with $p=0$. The green dots are exactly taken from Table 1-3 of \cite{Michalogiorgakis:2006jc}.}\label{Fig:QNscalar2}
\end{figure}  
 
 In Fig. \ref{Fig:QNscalar}  we give details for the hydrodynamic scalar QNM of the $\ell=2$ harmonic. We plot the real (imaginary) part of the frequency as a function of $r_+/L$ in the window of values where the frequency varies the most, namely $r_+/L<7$. The red point is the scalar normal mode frequency \eqref{SnormalM} for $p=0$. This plot complements the data in Table 1 and 2 of \cite{Michalogiorgakis:2006jc}  which is also represented as green dots in these plots (these are the QNM for $r_+/L=0.2,\, 1,\, 5$). For larger values of $r_+/L$ the real (imaginary) part keeps decreasing (increasing) monotonically  (for reference, for $r_+=100 L$ one has $\omega L\sim 1.732- 0.007\, i$ \cite{Michalogiorgakis:2006jc}). In the limit $r_+/L\to \infty$ it approaches the real value, $\omega L= 1.732$, dictated by \eqref{hydroS} which is plotted as a black dashed line for reference. In this  Fig. \ref{Fig:QNscalar}, the blue points are obtained solving numerically the Teukolsky system with BC \eqref{TheBCsSa0}, while the magenta points represent the numerical solution of the KI system with BC \eqref{KI:BCS}. It follows from  the map constructed in Section \ref{sec:mapKIvsTeuk} that these two curves have to overlap. This is indeed the case (the magenta points are drawn larger to be visible in the plot).
 
The evolution of the scalar QNM spectrum as the scalar hydrodynamic harmonic changes is displayed in Fig. \ref{Fig:QNscalar2} where we plot the curve(s) of Fig. \ref{Fig:QNscalar}, and in addition the hydrodynamic QNM of the next seven vector harmonics $\ell$. More concretely, from bottom to top (on the {\it Left} panel and on the right side of the {\it Right} panel) we have the harmonics $\ell=2,3,\cdots,10$. On the {\it Left} panel we normalize the real frequency to the respective $p=0$ normal mode frequency of AdS \eqref{SnormalM}. We see that the real part of the frequency spectrum is in a first approximation isospectral at each fixed $r_+/L$. From the {\it Right} panel, we conclude that the imaginary part of the frequency always has a minimum, unlike the vector case. Like in the vector QNM case, the inset plot shows that there is a window of $r_+/L$ around the global AdS case ($r_+/L=0$; red dot (0,0)) where the imaginary part of the spectrum is approximately flat increases, and this window increases as the harmonic $\ell$ grows. The green dots are exactly the values taken from Table 1,2 and 3 of \cite{Michalogiorgakis:2006jc}.

As a final remark, note that as discussed below \eqref{hKIvshTeuk}, the Teukolsky formalism describes only the harmonics with $\ell\geq |s|=2$. So, it misses the Kodama-Ishibashi vector mode with $\ell_{\rm \bf v}=1$ and scalar mode with $\ell_{\rm \bf s}=0$. The QNM spectrum of these KI modes is very special because it only contains a zero-mode, i.e. a mode with zero frequency. The scalar zero-mode produces  a shift in the mass of the solution, while the vector zero-mode generates angular momentum (thus connecting perturbatively global AdS-Schwarzschild to Kerr-AdS) \cite{Kodama:2003jz,Dias:2013hn}. The Teukolsky formulation is blind to these modes that generate deformations in the  conserved charges \cite{Wald,WaldL0L1}.

%%%%%%%%%%%%%%%%%%%%%%%%%%%%%%%%%%%%%%%%%%%%%%%%%%%
%%%SECTION:ACKNOWLEDGMENTS %%%
\section*{Acknowledgments}
It is a pleasure to warmly thank Gary Horowitz, Don Marolf and Harvey Reall for helpful discussions. OD thanks the Yukawa Institute for Theoretical Physics (YITP) at Kyoto University, where part of this work was completed during the YITP-T-11-08 programme  ``Recent advances in numerical and analytical methods for black hole dynamics", and the  participants of the workshops ``Iberian Strings 2013", Lisbon (Portugal), ``Holography, gauge theory and black holes", Amsterdam (Netherlands), ``XVIII IFT Xmas Workshop", Madrid (Spain),  ``The Holographic Way: String Theory, Gauge Theory and Black Holes", Nordita (Sweden), ``Spanish Relativity Meeting in Portugal",  ``Exploring AdS-CFT Dualities in Dynamical Settings", Perimeter Institute (Canada), and ``Numerical Relativity and High Energy Physics", Madeira (Portugal) for discussions.  JS acknowledges support from NSF Grant No. PHY12-05500. 

%%%%%%%%%%%%%%%%%%%%%%%%%%%%%%%%%%%%%%%%%%%%%%%%%%%%
%%%%%%%%%%%%%%%%%%%%%%%%%%%%%%%%%%%%%%%%%%%%%%%%%%%%
%%% BEGINNING OF APPENDIX %%%
\begin{appendix}
%%%%%%%%%%%%%%%%%%%%%%%%%%%%%%%%%%%%%%%%%%
\section{An overview of perturbations in Kerr(-AdS) \label{sec:completeness}}
%%%%%%%%%%%

The Teukolsky solutions $\delta\Psi_0$ and $\delta\Psi_4$, the Starobinski-Teukolsky identities, and the Hertz map that constructs the associated metric perturbations provide the {\it complete} information about the {\it most general} metric perturbation of the Kerr(-AdS) black hole \cite{Teukolsky:1973ha,Chandra1,Chandra2} (the only exception being the ``$\ell=0,1$" modes that simply add mass or angular momentum to the background \cite{WaldL0L1}; onwards we omit these exceptional modes from our discussion).

In this appendix we provide a brief historical overview of the studies that culminated with the above conclusion. In this discussion we assume the background to be Petrov type D and we highlight some facts that are sometimes not duly appreciated.

In the Newman-Penrose formalism, the fundamental gravitational variables are the 4 components of the NP tetrad basis ${\bf e}_{a}=\{\bm{\ell},\bm{n},\bm{m},\bm{\overline{m}}\}$, the 12 complex  NP spin coeficients \eqref{NPspincoef}  and the 5 complex NP Weyl scalars $\{\Psi_0,\cdots \Psi_4\}$. These variables are governed by a set of three systems of equations, namely, the Bianchi identities, the Ricci identities, and the commutation relations for the basis vectors. The metric is determined once we fix the  tetrad basis by
\begin{equation}\label{gNP}
g_{\mu\nu}=-2 \bm{\ell}_{(\mu}\bm{n}_{\nu)}+ 2 \bm{m}_{(\mu}\bm{\overline{m}}_{\nu)}\,.
\end{equation}  

The most general perturbation of this system requires determining 10+24+16=50 real functions to specify the perturbations of the 5 complex Weyl scalars, 12 complex spin coeficients and the 16 matrix components $A_a^b$ that describe the deformations of the tetrad via $\delta {\bf e}_{a}=A_a^b {\bf e}_{b}$.

In the Kerr(-AdS) background, this general perturbation system divides into two sectors \cite{Chandra1,Chandra2}:
 \begin{eqnarray}
&&  \:  \: I)  \quad \delta\Psi_0,\:\delta\Psi_1, \: \delta\Psi_3,  \: \delta\Psi_4, \:  \delta \kappa,  \: \delta \sigma,  \: \delta \lambda,  \: \delta\nu\,; \label{pertNP1} \\ 
&& II) \quad \delta\Psi_2,  \: \delta\alpha,  \: \delta\beta,  \: \delta\epsilon,  \: \delta\gamma,  \: \delta\pi, \:  \delta\rho,  \: \delta\mu,  \: \delta\tau,  \: \delta\bm{\ell},  \: \delta\bm{n},  \: \delta\bm{m},  \: \delta\bm{\overline{m}}\,.  \label{pertNP2}
\end{eqnarray}  
The first family describes perturbations of those variables that vanish in the Kerr(-AdS) background because it is a Petrov type D geometry. The second involves all other quantities that are not required to vanish in such a Petrov background.\footnote{This includes the spin coefficient $\epsilon$ that would vanish if we worked with an appropriate tetrad basis, but not otherwise. In particular it is not required to vanish by the algebraically special character of the spacetime.} A remarkable property is that these two sectors of perturbations ``almost decouple" in the sense that we can solve the perturbation sector $I)$ without solving the NP equations involving the perturbations of sector $II)$. The solutions of sector $I)$ are however a prerequisite to then search for the solutions of sector $II)$ \cite{Chandra1,Chandra2}.

Teukolsky \cite{Teukolsky:1973ha}, and later Chandrasekhar \cite{Chandra1}  following an independent computation,  found the solutions of the perturbation sector $I)$. One just needs to solve the Teukolsky master equation that gives the solution for $\delta\Psi_0$, say. The Starobinski-Teukolsky identities fix the relative normalization between these two variables \cite{Staro,StaroChuri,Teukolsky:1974yv,Chandra1,Chandra2}. These scalars $\delta\Psi_0$ and  $\delta\Psi_4$ are gauge invariant, i.e. invariant both under infinitesimal coordinate transformations and infinitesimal changes of NP basis. We can then set $\delta\Psi_1=0=\delta\Psi_3$ by an infinitesimal rotation of the tetrad basis. Finally, the perturbations of the spin coefficients $\delta \kappa, \delta \sigma,\delta \lambda,\delta\nu$ are obtained by applying differential operators to $\delta\Psi_0$ and  $\delta\Psi_4$. So the information on the linear perturbation of our system $I)$ is encoded in the gauge invariant variables  $\delta\Psi_0$ and  $\delta\Psi_4$. (This is not the full story concerning perturbation sector $I)$; we will come back and complete it in the end of this Appendix).

The most general perturbation problem is however not yet solved since we still need to find the solutions for the perturbation sector $II)$ in \eqref{pertNP2}. In a {\it tour de force} computation that requires starting with sector $I)$ solutions,  Chandrasekhar did a direct and complete integration of the remaining linearized NP equations to find the sector $II)$ solutions  \eqref{pertNP2}  \cite{Chandra2}. Remarkably, in the end of the day, all sector $II)$ perturbations are  determined also only as a function of $\delta\Psi_0,\:\delta\Psi_4$. This justifies the statement that the Teukolsky master equations and the Starobinski-Teukolsky identities encode the complete information about  general perturbations of the Kerr(-AdS) black hole. Note in particular that with the knowledge of the basis vector perturbations we can also easily construct the perturbations of the metric components through the variation of \eqref{gNP} \cite{Chandra2}.  

An astonishing twist in this story is that we do not need the major effort of integrating the full system of NP equations to get what is often the most desired result, namely the metric perturbations $h_{ab}$. This was first realized by Cohen and Kegeles \cite{Cohen:1975,Cohen:1979}  and Chrzanowski  \cite{Chrzanowski:1975wv} who have assumed some {\it ad hoc}, but smart guessed, hypothesis to build the Hertz map. At this point in time, this map was a prescription to reconstruct the most general perturbations of the metric tensor {\it only} from the knowledge of the Teukolsky master solutions $\delta\Psi_0,\:\delta\Psi_4$, i.e. without requiring information on the variables \eqref{pertNP2} (in a Petrov type D background). An elegantly simple proof  of the  Hertz map construction was finally provided by Wald in \cite{Wald}. It promotes the Hertz construction from a prescription into a formal map. Keypoints in Wald's proof are:
1) the existence of the decoupled Teukolsky master equations \eqref{TeukPosSpin}-\eqref{TeukNegSpin} for  $\delta\Psi_0,\:\delta\Psi_4$;
2) the fact that the Teukolsky  operator for  $\delta\Psi_0$ is the adjoint of the one for $\delta\Psi_4$, which on the other hand is in the end of the day responsible for the existence of the Starobinsky-Teukolsky differential  identities \eqref{StaroTeukRad}-\eqref{StaroTeukAng};
3) the fact that  the equations \eqref{HertzMaster} defining the Hertz map are the adjoint of the original Teukolsky master equations \eqref{TeukPosSpin}-\eqref{TeukNegSpin}.
Ultimately all these properties are due to the algebraically special character of the background. 

The upshot of Wald's proof of the Hertz map prescription of \cite{Cohen:1975}-\cite{Cohen:1979} is that we can use this map \eqref{HertzIRG}-\eqref{HertzORG} to obtain the complete, most general, metric perturbation of the Kerr(-AdS) black hole (with the exception of the modes that change the mass and angular momentum), without needing to integrate the extra NP equations that would be necessary to find the solutions \eqref{pertNP2}. 
It is worth to look back and appreciate this result: {\it \`a priori} we had to find a total of 50 variables. However, in the end of the day we just need to solve the Teukolsky master equation for $\delta \Psi_0$ (the equation for $\delta \Psi_4$ is its adjoint and their relative normalization set by the ST identities) to get through the Hertz map the most general metric perturbation. An important corollary of this result is that the global AdS boundary conditions we find in Section \ref{sec:AsymGlobalAdS} apply to generic  perturbations of the Kerr-AdS black hole (with $\ell\geq 2$).

We now return to an issue that was left  without its complete discussion. As said above, once we have completed  the analysis of  \cite{Teukolsky:1973ha,Chandra1}  we find the solution of perturbation sector $I)$. Strickly speaking there is however some residual incompleteness in this solution since  \cite{Chandra1,Chandra2}: 1) at this point we just know the absolute value but not the real and imaginary parts of the angular Starobinski-Teukolsky constant $\mathcal{C}_{\rm st}$, and 2) there is still an unknown numerical factor needed to fully determine the perturbations  $\delta \lambda$,  and $\delta\nu$. Both these gaps in our knowledge are filled once we solve perturbation sector $II)$ via an integrability condition \cite{Chandra1,Chandra2} (also reviewed in sections 82 to 95 of chapter 9 of the textbook \cite{ChandrasekharBook}).
To our knowledge, the analogous computation of  \cite{Chandra2} that determines this information in the Kerr-AdS case was never done, and it would be interesting to undergo this task. However, to determine the asymptotically global AdS BCs of Section \ref{sec:AsymGlobalAdS} we do not need this knowledge at all. Moreover, we do not need the explicit expression for $\mathcal{C}_{\rm st}$ to construct the map between the Kodama-Ishibashi and the $a=0$ Teukolsky formalisms of Section \ref{sec:mapKIvsTeuk}. 
Nevertheless, and for completeness,  in the main text we conjectured  the expression for $\mathcal{C}_{\rm {st}}$ to be the solution of $|\mathcal{C}_{\rm{st}}|^2$ as given in \eqref{STconstCf} that reduces to the asymptotically flat expression of \cite{Chandra2} when $L\to \infty$. This expression is written in \eqref{STconstCf} or \eqref{a0:STconst} when $a\to 0$. This is a reasonable expectation but  it would nevertheless be important to confirm this expression with a computation similar to the one done in \cite{Chandra2}. This would close the Kerr-AdS linear gravitational perturbation programme.

It is believed that Einstein's equation is not a special system of coupled PDEs. On the other hand, when these equations are linearized around a Petrov type D background and written in the Newman-Penrose formalism it is astonishing to find how special the linearized PDE system is.

%%%%%%%%%%%%%%%%%%%%%%%%%%%%%%%%%%%%%%%%%%%%%%%%%%%%
\end{appendix}

%%%%%%%%%%%%%%%%%%%%%%%%%%%%%%%%%%%%%%%%%%%%%%%%%%%%
%%%%%%%%%%%%%%%%%%%%%%%%%%%%%%%%%%%%%%%%%%%%%%%%%%%%
%bibliography%

%%%%%%%%%%%%%%%%%%%%%%%%%%%%%%%%%%%%%%%%%%%%%%%%%%%%
\end{document}